\pdfoutput=1
\documentclass[a4paper,UKenglish,numberwithinsect]{lipics-v2021}

\usepackage{mathtools,todonotes}
\usepackage{booktabs,enumitem}
\usepackage{tabularx}
\usepackage{placeins}
\usepackage{tikz,xcolor}
\usetikzlibrary{arrows.meta,positioning,fit,calc,decorations.pathreplacing}

\newcommand{\Av}{\operatorname{Av}}
\newcommand{\mm}[1]{\lVert #1 \rVert}

\renewcommand{\st}{\operatorname{st}}

\newcommand{\LDS}{\operatorname{LDS}}
\newcommand{\ins}{\operatorname{ins}}

\newcommand{\ExtendMax}
{\textnormal{\textsc{ExtendMax}}}
\newcommand{\FPCSP}{\textnormal{\textsc{FPCSP}}}
\newcommand{\EnumAvoid}{\textnormal{\textsc{EnumAvoid}}}
\newcommand{\EnumExt}{\textnormal{\textsc{EnumExt}}}

\newcommand{\B}{\lVert B\rVert}
\newcommand{\Bstate}{B_{\mathrm{state}}}
\newcommand{\Bfinal}{B_{\mathrm{final}}}

\newcommand{\Pneq}[1]{P^{\neq}_{#1}}
\newcommand{\Pmono}[1]{P^{\mathrm{mono}}_{#1}}
\newcommand{\Aedge}{A_{\mathrm{eq}}}
\newcommand{\Anae}{A_{\mathrm{nae}}}

\DeclareMathOperator{\NP}{NP}

\DeclareMathOperator{\bin}{bin}

\newcommand{\NPhard}{\texorpdfstring{\mbox{NP-hard}}{NP-hard}}
\newcommand{\NPcomplete}{\texorpdfstring{\mbox{NP-complete}}{NP-complete}}
\newcommand{\NPhardness}{\texorpdfstring{\mbox{NP-hardness}}{NP-hardness}}
\newcommand{\NPcompleteness}{\texorpdfstring{\mbox{NP-completeness}}{NP-completeness}}

\newenvironment{permpic}[2][0.26]{%
  \begin{tikzpicture}[scale=#1,baseline={([yshift=-0.6ex]current bounding box.center)}]
  \draw[gray!16,very thin] (0.5,0.5) grid ({#2+0.5},{#2+0.5});
  \draw[gray!50] (0.5,0.5) rectangle ({#2+0.5},{#2+0.5});
}{\end{tikzpicture}}
\newcommand{\pts}[2][black]{%
  \foreach \v [count=\i] in {#2} {\fill[#1] (\i,\v) circle[radius=0.32];}}

\title{Enumeration Complexity of Permutation Classes}
\titlerunning{Enumeration Complexity of Permutation Classes}
\author{Manuel Bodirsky}{TU Dresden, Germany}{}{}{Funded by the European Union (project POCOCOP, ERC Synergy grant
No.~101071674). Views and opinions expressed are however those of the
author(s) only and do not necessarily reflect those of the European
Union or the European Research Council Executive Agency. Neither the
European Union nor the granting authority can be held responsible for them.}
\author{Mohamed Elsheikh}{TU Dresden, Germany}{}{}{}
\authorrunning{M. Bodirsky and M. Elsheikh}
\Copyright{Manuel Bodirsky and Mohamed Elsheikh}
\ccsdesc[500]{Theory of computation~Problems, reductions and completeness}
\ccsdesc[500]{Theory of computation~Enumeration}
\ccsdesc[300]{Mathematics of computing~Permutations and combinations}
\keywords{Permutation patterns, enumeration complexity, polynomial delay, extension problems, constraint satisfaction}
\hideLIPIcs
\makeatletter
\let\@oddfoot\@empty
\makeatother
\AtBeginDocument{\hypersetup{
  pdfauthor={Manuel Bodirsky and Mohamed Elsheikh},
  pdftitle={Enumeration Complexity of Permutation Classes},
  pdfsubject={}
}}

\begin{document}
\nolinenumbers
\maketitle
\begin{abstract}
We study classes of permutations that are defined by avoidance of forbidden patterns. For each fixed finite set $B$ of forbidden patterns, we study the \emph{avoider enumeration problem}: given an integer $n$, output all $B$-avoiding permutations of length $n$. 
Avoider enumeration is output-polynomial for every fixed $B$. 
However, our main notion of efficiency is polynomial delay. We use the insert-maximum tree as a search tree for enumeration, in which permutations are built by repeatedly inserting a new maximum element while preserving pattern avoidance. This leads to the \emph{insert-maximum extension problem}, which asks whether a given permutation has an avoiding descendant at a specified depth. 
An algorithm for this problem can be used to prune branches in the search tree that cannot reach the target length. Hence, if this problem is in P, we obtain polynomial-delay enumeration.

To investigate the complexity of the insert-maximum extension problem, we introduce and study \emph{pockets}, the maximal finite subtrees of the insert-maximum tree. For fixed $B$, if there is a polynomial $p$
such that every pocket rooted at $\rho$
in its insert-maximum tree has at most $p(|\rho|)$ nodes,
then the extension problem is solvable in
polynomial time and enumeration has
polynomial delay. In particular, it suffices that every pocket has depth at most a constant depending only
on $B$. We prove such a bound if the class
is well-quasi-ordered, has a regular insertion
encoding, or if $B$ omits one of the three
maximum-position types: leftmost, interior,
or rightmost.
We also prove that the extension problem is in NP for every fixed $B$, even if the target depth is encoded in binary.

On the negative side, we construct a fixed set of $55$ forbidden patterns for which extension is NP-complete. Consequently, enumerating valid extensions from a given starting permutation cannot have polynomial delay unless P=NP. This also answers a question of Hanika by showing that the forbidden permutation constraint satisfaction problem can be NP-complete for a fixed forbidden set.
\end{abstract}

\paragraph*{Disclosure of AI assistance.}
ChatGPT with the GPT 5.6 Sol model was used during development of the fixed
forbidden set $\Bfinal$.  The interaction used \emph{Generalized Coloring of
Permutations} by Jel\'inek, Opler, and Valtr~\cite{JelinekOplerValtr2024}
as background and involved about $150$ prompts over about four days.
An initial suggested construction had $224$ forbidden patterns, the longest
of length $481$.  The suggestions to use increasing oscillations and simple
permutations came from ChatGPT. The routing idea came from the provided paper. The construction was subsequently checked
and simplified by the authors. 
The entire paper was thoroughly revised by the authors, and GPT 6 Astra and Claude Opus 5.5 were used for proofreading. The authors take full responsibility for the final text.

\section{Introduction}
\label{sec:introduction}
Permutation patterns
describe the relative order of selected entries in a
permutation.  A permutation \emph{contains} a pattern if some subsequence has the
same relative order as that pattern, and \emph{avoids} it otherwise.  For example,
$2413$ contains $132$ through the entries $2,4,3$, but avoids $321$.
Pattern avoidance provides
a common language for structural and counting questions about
permutations~\cite{Vatter2015}.

For a finite set of permutations $B$, let $\Av_n(B)$ be the permutations of length
$n$ that avoid every member of $B$.  %
We study the algorithmic task of enumerating all members of $\Av_n(B)$. 
Here \emph{enumeration} means outputting the objects (whereas \emph{counting}
means computing their number).
For fixed $B$, let $\EnumAvoid_B$ denote the problem of enumerating
all members of $\Av_n(B)$, without repetitions, where $n$ is given in unary.

An enumeration algorithm has \emph{polynomial delay} if the time before
the first output, between consecutive outputs, and between the last output
and termination is bounded by a polynomial in the input size.  
This also
requires polynomial-time termination when there is no output.  We write
$\mathrm{DelayP}$ for the class of enumeration problems admitting such an
algorithm.
An enumeration algorithm is \emph{output-polynomial} if its total running
time is bounded by a polynomial in the input size and the total encoded size
of all outputs.  We write $\operatorname{OutputP}$ for the corresponding
class~\cite{JohnsonYannakakisPapadimitriou1988,Strozecki2021}.

Successive insertions of new maxima give a natural search tree, which we call the  \emph{insert-maximum tree}. When the class is nonempty, the root is
the empty permutation.  The children of a permutation are obtained by
inserting the next maximum such that the resulting permutation is still in
the class.  Deleting the maximum gives each nonroot node a unique parent,
so level $n$ contains exactly $\Av_n(B)$, without repetitions.  Therefore, the required permutations can be enumerated with depth-first search.  The difficulty is
that a finite branch may contain many nodes but no permutation of the target
length.

To guide the traversal, we consider an auxiliary decision problem.  Fix a
finite forbidden set $B$.  Given a permutation $\rho$ and a nonnegative
integer $k$, the problem $\ExtendMax_B$ asks whether $\rho$ has a
$B$-avoiding extension obtained by inserting $k$ new maxima.  Equivalently,
it asks whether the subtree rooted at $\rho$ reaches depth $k$.  If this
problem is solvable in polynomial time, then \emph{flashlight search} (a general method for turning an extension oracle into
an enumeration algorithm~\cite[Section~3.3]{Strozecki2021}) can discard every
branch that contains no output and thereby enumerate the target level with
polynomial delay. We use $\ExtendMax_B$ with the superscripts $\mathrm{un}$ and $\mathrm{bin}$ to specify whether the given $k \in {\mathbb N}$ is encoded in unary or in binary.

\subsection{Results}
\label{sect:results}
Every fixed finite set $B$ of forbidden permutations admits
output-polynomial avoider enumeration. We give sufficient conditions on $B$ 
for polynomial-time extension and polynomial-delay enumeration.  For
a nonempty set of permutations $B$, write
$\mm{B}:=\max_{\beta\in B}|\beta|$.  A node is \emph{viable} if it has extensions
of arbitrary depth, and \emph{mortal} otherwise.  A \emph{pocket} is a maximal
finite subtree.

\begin{theorem}
\label{thm:main-positive}
For every finite set of permutations $B$, the following hold.
\begin{enumerate}
\item $\EnumAvoid_B\in\operatorname{OutputP}$.
\item $\ExtendMax_B^{\mathrm{bin}}$ belongs to $\NP$. For nonempty $B$
      containing only nonempty patterns, every mortal permutation $\rho$
      has depth at most $(\mm{B}-1)^2(|\rho|+1)$.
\item If there is a polynomial $p$ such
that every pocket rooted at $\rho$
has at most $p(|\rho|)$ nodes, then $\ExtendMax_B^{\mathrm{un}}$ and
      $\ExtendMax_B^{\mathrm{bin}}$ belong to $\mathrm P$, and there is a polynomial delay algorithm for the extension
      enumeration problem.
\item If $\Av(B)$ is well-quasi-ordered, if it has a regular
      insertion encoding, or if $B$ consists of nonempty patterns and omits
      one of the leftmost, interior, or rightmost maximum-position types,
      then there is a constant $D$ such that every pocket has depth at most
      $D$.  Consequently, every pocket rooted at $\rho$ has size bounded by
      a polynomial in $|\rho|$, as required in the preceding item.
\end{enumerate}
\end{theorem}
A partial order is \emph{well-quasi-ordered} if it contains neither an infinite strictly descending chain nor an infinite antichain. Since the poset of finite permutations under pattern containment has no infinite strictly descending chains, a permutation class is well-quasi-ordered if and only if it contains no infinite antichain. This concept has been studied e.g. in~\cite{BrignallVatter2022}. \emph{Regular insertion encodings} have been introduced in~\cite[Section~4]{AlbertLintonRuskuc2005}.

Our main hardness result shows that efficient extension cannot be expected
for every fixed basis.  The result holds even when the forbidden set is
fixed: only the starting permutation and the target depth are part of the
input.

\begin{theorem}
\label{thm:main-hardness}
There exists a set $\Bfinal$ of $55$ permutations, each
of length at most $191$, such that
$\ExtendMax_{\Bfinal}^{\mathrm{un}}$ and
$\ExtendMax_{\Bfinal}^{\mathrm{bin}}$ are NP-complete.
\end{theorem}

We prove NP-hardness by reduction from monotone NAE-3SAT.  In the constructed starting
permutation, each Boolean state is represented by a pair of \emph{insertion sites};
choosing one of the two sites encodes a bit.  Fixed forbidden patterns ensure
that an extension of maximum possible depth chooses exactly one site from
every pair.
Further patterns enforce equality along wires and the not-all-equal
condition on clause triples.  Increasing oscillations and simple marker
permutations make these local constraints recognizable by ordinary pattern
containment.  A substantial part of the proof excludes unintended
occurrences of the forbidden patterns.

For a fixed finite set $B$ of forbidden permutations, Hanika introduced the \emph{forbidden permutation constraint satisfaction problem}
$\FPCSP(B)$, where the input consists of two acyclic relations on a common finite set, and the task is to decide whether these relations
can be completed to linear orders whose induced permutation avoids the fixed
forbidden set $B$~\cite[Problem~4.1]{Hanika2015}.  Hanika asked whether
there exists a fixed finite set $B$ of forbidden patterns for which
$\FPCSP(B)$ is NP-complete.  Since insert-maximum extension is a
restriction of this problem, Theorem~\ref{thm:main-hardness} answers his
question affirmatively: $\FPCSP(\Bfinal)$ is NP-complete.

The hardness conclusion does not transfer to avoider enumeration.  In
particular, we do not prove that enumerating $\Av_n(\Bfinal)$ lacks a
polynomial-delay algorithm.  The distinction between an empty start and a
supplied start is essential.

\subsection{Related work}

The insert-maximum tree is also called the \emph{pattern-avoidance
tree}~\cite[Section~1]{Vatter2006}.  Kuszmaul uses the same insertion rule
in an output-sensitive algorithm generating avoiders of all lengths at most
$n$~\cite[Algorithm~2]{Kuszmaul2018}.  Its runtime bound involves the
total size of these levels and does not directly control the delay at a single
target level.  Gray-code constructions give efficient generation for
several particular classes~\cite{DukesEtAl2008,DoTranVajnovszki2019}.
Hartung, Hoang, M\"utze, and Williams develop a broader generation framework
based on permutation languages~\cite{HartungEtAl2022}. 


\section{The insert-maximum tree and enumeration}
\label{sec:permutation-background}
\label{sec:insert-maximum-tree}

We first fix the notation and the input conventions.  The set $B$ of forbidden permutations is
fixed throughout the paper.  All algorithmic constants may depend on it.

A permutation $\pi=\pi_1\cdots\pi_n$ of length
$|\pi|=n$ contains each element of
$[n]=\{1,\ldots,n\}$ exactly once.
We call $\pi_i$ the entry at position $i$. The empty permutation is
denoted by $\varepsilon$.
For distinct numbers $a_1,\ldots,a_m$,
their \emph{standardization}
$\st(a_1,\ldots,a_m)$ is the unique
permutation with the same relative order.
A set of entries
$\{\pi_{i_1},\ldots,\pi_{i_m}\}$,
where $i_1<\cdots<i_m$, is an
\emph{occurrence} of a permutation $\tau$
in $\pi$ if
$
  \st(\pi_{i_1},\ldots,\pi_{i_m})=\tau.
$
We call $\tau$ a \emph{pattern} of $\pi$
and write $\tau\leq\pi$ if such an
occurrence exists. The entries of an
occurrence \emph{induce} $\tau$. Define
$
  \Av(B)=\{\pi:\beta\not\leq\pi
  \text{ for all }\beta\in B\}.
$
A basis is an antichain of minimal excluded permutations.  A finite forbidden
set need not be a basis, but deleting its redundant members does not change
the class.  For nonempty $B$, write
$\B:=\max_{\beta\in B}|\beta|$. 

A permutation of length $n$ has \emph{gaps} $0,\ldots,n$, which are the positions before and after the entries of the permutation; gap $g \in \{1,\dots,n-1\}$ is between the $g$-th and the $g+1$-st entry.
Let $\ins_g(\pi)$ be the permutation obtained from $\pi$ by inserting $n+1$ into gap $g$.
A gap is \emph{active} if this permutation belongs to $\Av(B)$ and
\emph{inactive} otherwise.  The children of $\pi$ in the insert-maximum
tree are the permutations of the form $\ins_g(\pi)$ for the active gaps $g$ of $\pi$.

An \emph{extension} of $\rho$ is a permutation $\sigma\in\Av(B)$ such that
deleting its $|\sigma|-|\rho|$ largest entries leaves $\rho$.  We write
$\rho\rightsquigarrow\sigma$.  Its \emph{depth over $\rho$} is
$|\sigma|-|\rho|$.  Hereditary avoidance ensures that the unique sequence
of successive maximum deletions from $\sigma$ to $\rho$ stays in the class.
Put
\[
 d_B(\rho)=\sup\{\,|\sigma|-|\rho|:\rho\rightsquigarrow\sigma\,\}.
\]
For $\rho\in\Av(B)$ this belongs to $\mathbb Z_{\geq0}\cup\{\infty\}$.
Given $\rho\rightsquigarrow\sigma$, we call the entries of $\sigma$
inherited from $\rho$ the \emph{root entries with respect to $\rho$}, or
simply the \emph{root entries} when $\rho$ is clear. The \emph{root gap} of an
inserted entry $x$ is the number
of root entries appearing to the left of $x$ in $\sigma$.

The input to $\ExtendMax_B$ is $(\rho,k)$, and the answer is Yes exactly
when $\rho$ has an extension of depth $k$.  
For enumeration, $k$ is encoded in unary.  
This is natural
because an extension at depth $k$ has length $|\rho|+k$, so with a binary
target even writing one output may require time exponential in the input
length.
On input $(\rho,1^k)$,
the extension enumeration problem $\EnumExt_B$ outputs all extensions of $\rho$ at depth $k$,
without repetitions.  On input $1^n$, the problem $\EnumAvoid_B$ is the
special case $\rho=\varepsilon$ and $k=n$.

\begin{proposition}
\label{prop:existence-enumeration-equivalence}
For every fixed finite set of permutations $B$,
\[
 \ExtendMax_B^{\mathrm{un}}\in\mathrm P
 \quad\Longleftrightarrow\quad
 \EnumExt_B\in\mathrm{DelayP}
 \quad\Longrightarrow\quad
 \EnumAvoid_B\in\mathrm{DelayP}.
\]
\end{proposition}

\begin{proof}
Suppose that $D_B$
is an algorithm that decides whether a given permutation $\tau$ has an
extension of depth $j$ in time polynomial in $|\tau|+j$.
On input $(\rho,1^k)$, 
first execute $D_B(\rho,k)$ and
terminate if the answer is No. Otherwise, traverse the
subtree of the insert-maximum tree rooted at $\rho$ in
depth-first order, stopping at level $n:=|\rho|+k$.
For each child $\sigma$ of a visited permutation, descend
to $\sigma$ only if $D_B(\sigma,n-|\sigma|)$ returns Yes.
At level $n$, output the permutation.

The oracle tests ensure that only branches leading to
the target level are explored. Since every extension
has a unique path from $\rho$, every required extension
is output exactly once. Between consecutive outputs,
the search backtracks and descends through at most
$2k$ levels, considering at most $n+1$ children at
each level. Since each oracle call takes polynomial
time in $n$, the delay is polynomial. The same holds
before the first output and after the last.

Conversely, a polynomial-delay enumerator decides
extension existence in polynomial time by running
until its first output or termination.

Finally, setting $\rho=\varepsilon$ gives
$\EnumAvoid_B\in\mathrm{DelayP}$.
\end{proof}

\section{Output-polynomial avoider enumeration}
\label{sec:outputp}
An enumeration problem
belongs to $\operatorname{OutputP}$ if its total running time is polynomial
in the input size and the total size of the output.

\begin{theorem}
\label{thm:outputp-all-fixed-bases}
For every fixed finite set $B$ of permutations,
$\EnumAvoid_B\in\operatorname{OutputP}$.
\end{theorem}

\begin{proof}
Write $a_j:=|\Av_j(B)|$ and $A_{<n}:=\sum_{j=0}^{n-1}a_j$.
If $B = \emptyset$ or $\varepsilon\in B$ then the statement is trivial, so let us assume that this is not the case. 
We first claim that $A_{<n}$ is bounded by a polynomial in $n$ and $a_n$.
The Kaiser--Klazar growth gap says that a permutation class either has
polynomial growth or satisfies $a_n\geq\lambda^n$ for all sufficiently
large $n$ and some $\lambda>1$~\cite[Theorem~3.4]{KaiserKlazar2003}.
In the first case the claim is immediate.  In the second case,
Marcus--Tardos gives a constant $M>1$ with $a_j\leq M^j$ for all
$j$~\cite[Corollary~2]{MarcusTardos2004}.  Choose an integer $r$ with
$M\leq\lambda^r$.  For all sufficiently large $n$,
\[
 A_{<n}\leq\sum_{j=0}^{n-1}M^j
 \leq \frac{M^n}{M-1}
 \leq \frac{a_n^r}{M-1}.
\]
Increasing the polynomial to cover the finitely many remaining values of
$n$ proves the claim.

Now traverse the insert-maximum tree depth first from $\varepsilon$, stopping
at level $n$.  The traversal visits exactly the $A_{<n}$ avoiders below the
target level and outputs every member of $\Av_n(B)$ once.  For fixed $B$,
avoidance of $B$ can be tested in linear time: apply the algorithm of
Guillemot and Marx to each fixed pattern in $B$~\cite[Theorem~1.1]{GuillemotMarx2014}.
Each internal node has at most $n$ candidate children, each of length at most
$n$, so the total time is
\[
 O\!\left(n^2A_{<n}+n a_n\right).
\]
By the claim, this is polynomial in the input size and the total output size
$n a_n$.
\end{proof}

\section{A bound on finite extension depth}
\label{sec:depth}
In this section, we prove that $\ExtendMax^{\bin}_B$ belongs 
to $\NP$. If $(\rho,k)$ 
is an input to this problem, then we may not simply guess an extension of $\rho$ at depth $k$ in $\Av(B)$ non-deterministically, because
$k$ 
 might be exponential
 in the input size.
 However, we can exploit 
that a sufficiently
long extension contains a monotone subsequence that can be lengthened indefinitely.
We will use the following result of Erd\H{o}s--Szekeres~\cite[p.~468]{erdos-szekeres-1935}.

\begin{theorem}
\label{thm:erdos-szekeres}
Every sequence of $(u-1)(v-1)+1$ distinct numbers contains an increasing
subsequence of length $u$ or a decreasing subsequence of length $v$.
\end{theorem}

\begin{lemma}[Infinite-depth criterion]
\label{lem:infinite-depth-criterion}
Let $B$ be a nonempty finite set of nonempty permutations and let $\rho\in\Av(B)$ have length $n$.
Then the following are equivalent:
\begin{enumerate}[label=$(\roman*)$]
\item $d_B(\rho)=\infty$;
\item inserting $\mm{B}$ new maxima in
      monotone order into some gap of $\rho$ produces a permutation in $\Av(B)$;
\item $d_B(\rho)\geq (\mm{B}-1)^2(n+1)+1$.
\end{enumerate}
\end{lemma}
\begin{proof}
The implication (i)$\Rightarrow$(iii) is immediate.  Suppose (iii) holds.
Distribute the inserted entries of an extension of that depth among the
$n+1$ root gaps.  Some gap contains at least $(\mm{B}-1)^2+1$ inserted entries.
Theorem~\ref{thm:erdos-szekeres} gives a monotone subsequence of length $\mm{B}$
among them.  Delete all other inserted entries and standardize. The remaining new maxima form a monotone
subsequence in one root gap, proving (ii).

For (ii)$\Rightarrow$(i), insert $\ell\geq\mm{B}$
new maxima in monotone order into the same gap of $\rho$.
If the resulting permutation contained $\beta\in B$,
that occurrence would use at most
$|\beta|\leq\mm{B}$ inserted entries.
Since all inserted entries lie in the same gap
and are monotone, the same occurrence could
be formed using the $\mm{B}$ inserted entries from (ii), contradicting avoidance.
Arbitrarily many new maxima can be inserted
successively in this way, so $d_B(\rho)=\infty$.
\end{proof}

For fixed $B$, avoidance is polynomial-time testable by checking each
possible occurrence of each forbidden pattern.  In fact it is linear-time
testable by fixed-pattern matching~\cite{GuillemotMarx2014}, but the
elementary polynomial bound suffices here.  Consequently, viability is
polynomial-time decidable: test the $2(n+1)$ monotone extensions in
Lemma~\ref{lem:infinite-depth-criterion}.

\begin{corollary}
\label{thm:depthlimit}
Every mortal permutation $\rho$ has depth at most
$(\mm{B}-1)^2(|\rho|+1)$, under the assumptions of
Lemma~\ref{lem:infinite-depth-criterion}.
\end{corollary}

\begin{theorem}
\label{thm:extendmax-in-np}
\label{cor:fixed-extendmax-unary-in-np}
For every finite set $B$ of permutations, $\ExtendMax^{\bin}_B$ belongs
to $\NP$.
\end{theorem}
\begin{proof}
If $B=\varnothing$, every permutation has extensions of every depth.  If
$\varepsilon\in B$, no avoiding permutation exists. So suppose that $B \neq \varnothing$ does not contain $\varepsilon$. 
Reject if $\rho \notin \Av(B)$. Put $n:=|\rho|$ and 
$H:=(\mm{B}-1)^2(n+1)+1$.  If $k<H$, guess an extension of length $n+k$ and
check avoidance and the restriction to the $n$ smallest entries.  Its
length is in $O(n+1)$, because $B$ is fixed.  If $k\geq H$, Lemma~\ref{lem:infinite-depth-criterion} reduces the question to viability, which is polynomial-time
decidable. 
\end{proof}

\section{Pockets and polynomial delay}
\label{sec:pockets}
The finite subtrees of the insert-maximum tree that contain no output are the possible source
of long delays.  

\begin{definition}
A \emph{pocket} is the subtree of the  insert-maximum tree rooted at a mortal
node whose parent, if it exists, is viable.  
The \emph{size} of a pocket is its number of nodes, including
its root, and its \emph{depth} is the number of edges of the longest path starting at the root.
\end{definition}

So pockets are exactly the maximal finite subtrees.  The insert-maximum tree is
finitely branching, so a subtree of finite depth is finite.  Every
descendant of a mortal node is mortal, and every mortal node lies in a
unique pocket, obtained by following parents until mortality first begins.

\begin{theorem}
\label{thm:poly-pockets-extension}
Suppose a nondecreasing polynomial $q$ bounds the size of every pocket
rooted at $\rho$ by $q(|\rho|)$.  Then both target encodings of
$\ExtendMax_B$ are polynomial-time solvable, and both 
$\EnumExt_B$ and $\EnumAvoid_B$ are in $\mathrm{DelayP}$.
\end{theorem}
\begin{proof}
The cases $B=\varnothing$ and $\varepsilon\in B$ are immediate.  On input
$(\rho,k)$, first check $\rho\in\Av(B)$.  Decide viability by the
infinite-depth criterion (Lemma~\ref{lem:infinite-depth-criterion}).  If $\rho$ is viable, accept for every $k$.
Otherwise its subtree lies in a pocket rooted at an ancestor $\alpha$.
Since $|\alpha|\leq|\rho|$, that pocket has at most $q(|\rho|)$ nodes.
Traverse the subtree of $\rho$ and accept precisely when depth $k$ is
reached.  Each visited permutation has length at most
$|\rho|+q(|\rho|)$, so constructing and checking all its candidate
children takes polynomial time.  There are polynomially many visited
nodes.  The algorithm works even for a binary target, since it never
iterates up to $k$ unless that depth is reached in the finite subtree.
The enumeration conclusions follow from
Proposition~\ref{prop:existence-enumeration-equivalence}.
\end{proof}

\begin{corollary}
\label{cor:bounded-pocket-depth-consequences}
A uniform bound on pocket depth implies all conclusions of
Theorem~\ref{thm:poly-pockets-extension}.
\end{corollary}
\begin{proof}
If each pocket has depth at most $D$, then a pocket rooted at length $m$
has at most $(D+1)(m+D+1)^D$ nodes.  For fixed $D$ this is a
nondecreasing polynomial in $m$.
\end{proof}

Corollary~\ref{thm:depthlimit} bounds the depth of every pocket by a linear
function of its root length.  This alone does not give a polynomial size
bound: the exponent in the counting argument in the proof of 
Corollary~\ref{cor:bounded-pocket-depth-consequences} 
would no longer
be constant.  The next section gives conditions that remove dependence on
the root length.

\section{Structural conditions for bounded pocket depth}
\label{sec:structural}

We give three sufficient conditions for bounded pocket depth.

\subsection{Well-quasi-ordered classes}
Recall from Section~\ref{sect:results} that a class of permutations is \emph{well-quasi-ordered} if it contains no infinite antichain
under pattern containment. 
For example, the class 
$\Av(\{2413,3142\})$ of all \emph{separable permutations} is well-quasi-ordered~\cite{AlbertAtkinson2005}.
A \emph{trap} is a mortal permutation minimal
under containment among the mortal members of the class.

\begin{lemma}
\label{thm:depth-monotonicity}
If $\pi\leq\sigma$ and both belong to $\Av(B)$, then
$d_B(\sigma)\leq d_B(\pi)$.
\end{lemma}
\begin{proof}
Fix an occurrence of $\pi$ in $\sigma$.  In an arbitrary extension of
$\sigma$, retain this occurrence and all inserted entries, deleting every
other root entry.  Standardization gives an extension of $\pi$ at the
same depth.  It avoids $B$ by heredity.  Every depth attainable from
$\sigma$ is therefore attainable from $\pi$.
\end{proof}

\begin{theorem}
\label{thm:wqo_bounded_pocket_depth}
If $\Av(B)$ is well-quasi-ordered, its pocket depth is uniformly bounded.
\end{theorem}
\begin{proof}
Distinct traps are incomparable by minimality, so the traps form an
antichain.  There are therefore only finitely many.  If there are none,
every node is viable.  Otherwise, let $D$ be the maximum of their finite
depths.  Every pocket root $\rho$ contains a trap $\tau$, and
Lemma~\ref{thm:depth-monotonicity} gives
$d_B(\rho)\leq d_B(\tau)\leq D$.
\end{proof}

This is a fixed-class result.  The proof supplies a constant depending on
$B$; it does not claim an algorithm computing that constant from an arbitrary
input basis.  The extension algorithm itself only needs the viability test
and traversal, not an explicit list of traps.

\subsection{Regular insertion encodings}
Insertion encodings record how a permutation is built by inserting its entries in increasing order. 
 A \emph{configuration}
 is a word containing the entries already inserted and zero or more \emph{slots}, denoted by $\diamond$. 
An \emph{insertion} replaces the $j$-th 
slot by $v$ ($f_j$), $v\diamond$ ($l_j$), $\diamond v$ ($r_j$), or $\diamond v\diamond$ ($m_j$) where $v$ is the next maximum. 
These choices $f_j$, $l_j$, $r_j$, $m_j$ record whether the new value fills the $j$-th slot, leaves space on the right, on the left, or on both sides.   
Starting from the single slot $\diamond$, the sequence of operations that ends at a permutation $\pi$ is called its \emph{insertion encoding}.
It can be viewed as a word over the infinite alphabet $\bigcup_{j \geq 1} \{f_j,l_j,r_j,m_j\}$. For a permutation class ${\mathcal C}$, we write $L({\mathcal C})$ for the set of all insertion encodings of permutations in ${\mathcal C}$. For some classes ${\mathcal C}$, only finitely many symbols of the alphabet appear, and sometimes $L({\mathcal C})$ is even a regular language. In this case we say that ${\mathcal C}$ has a \emph{regular insertion encoding}.
An example of such a class is 
$\Av(321, 3412)$ (see, e.g.,~\cite{AlbertLintonRuskuc2005}; it is not well-quasi-ordered~\cite[Lemma 4.5 and its proof]{AtkinsonMurphyRuskuc2002}). 
We need the following result from~\cite[Section~4]{AlbertLintonRuskuc2005}.

\begin{theorem}\label{thm:k-bounded}
    Let $B$ be a finite set of permutations. Then 
    $\Av(B)$ has a regular insertion encoding if and only if there exists a $k \in {\mathbb N}$ such that every configuration that appears in the insertion encoding of a permutation from $\Av(B)$ contains at most $k$ slots. 
\end{theorem}

\begin{theorem}
\label{thm:regular-bounded-pocket-depth}
Let $B$ be a finite nonempty set of nonempty patterns.  If $\Av(B)$ has a
regular insertion encoding,
then its pocket depth is uniformly bounded. 
\end{theorem}
\begin{proof}
Let $k$ be the number from Theorem~\ref{thm:k-bounded}. 
Let $\rho \in \Av(B)$ be mortal and let $\sigma$ be an extension of $\rho$. In the insertion
encoding of $\sigma$, consider the configuration just after the entries of
$\rho$ have been inserted.  Its slots correspond exactly to the root gaps
used by the inserted entries of $\sigma$.  Thus at most $k$ root gaps are
used.  No such gap can contain more than $(\mm{B}-1)^2$ inserted entries:
otherwise Erd\H{o}s--Szekeres (Theorem~\ref{thm:erdos-szekeres}) and the infinite-depth criterion (Lemma~\ref{lem:infinite-depth-criterion}) would make
$\rho$ viable. Thus, the total extension depth is at most $k(\mm{B}-1)^2$.
\end{proof}

\subsection{Maximum-position types}
Call a nonempty pattern an \emph{$L$-pattern} if its maximum is leftmost, an
\emph{$R$-pattern} if its maximum is rightmost, and an \emph{$M$-pattern} if its maximum
is at neither endpoint.  The singleton pattern has both endpoint types. In this section we will prove the following. 

\begin{theorem}
\label{thm:maximum-types}
Let $B$ be a finite set of nonempty permutations.  
If $B$ has no $L$-pattern or
no $R$-pattern, then $\Av(B)$ has no pockets.  If $B$ is nonempty and has no
$M$-pattern, then every pocket of $\Av(B)$ has depth at most $(\mm{B}-1)^2$.
\end{theorem}

For a permutation $\tau$ and $0\leq i\leq|\tau|$, let $\tau^{-i}$ be the
permutation obtained by deleting its $i$ largest entries. For $\tau \in \Av(B)$ and an active gap
$g$ of $\tau$, the \emph{increasing run at $g$} 
inserts the maximum at gap $g$, and then  
repeatedly inserts the next maximum
immediately to the right of the preceding insertion, as long as the result stays in $\Av(B)$.
The \emph{decreasing
run} is defined analogously, but uses the left side instead. The \emph{length} of a run is the number of insertions, and may be infinite.

\begin{lemma}[One-sided blocking]
\label{lem:one-sided-blocking}
Assume that $B$ is nonempty and has no $M$-pattern.  If the increasing run
at an active gap $g$ of a permutation $\pi\in\Av(B)$ has finite length, then there is an
$R$-pattern $\beta\in B$ and an integer $2\leq p\leq|\beta|\leq \mm{B}$ such
that 
\begin{itemize}
    \item $\beta^{-p}$ occurs entirely to the left of $g$,
    \item the last $p$ entries of $\beta$ are its $p$ largest entries, appearing in increasing order, 
\item the increasing run at $g$ has length
$p-1$. 
\end{itemize}
Symmetrically, a finite decreasing run yields an $L$-pattern whose first
$q$ entries are its $q$ largest entries, in decreasing order, and whose remaining entries
occur to the right of $g$.
\end{lemma}

\begin{proof}
Let the increasing run have length $k$, with inserted entries
$x_1<\cdots<x_k$, and let $x_{k+1}$ be the first unsuccessful insertion.
Note that $k \geq 1$ since $g$ is active.
Choose an occurrence of some $\beta\in B$.  It uses $x_{k+1}$ as
its maximum.  Since $B$ has no $M$-pattern, this maximum is at an endpoint.
It cannot be the left endpoint: all other entries of such an occurrence
would lie to its right, and replacing $x_{k+1}$ by $x_k$ would reproduce
the occurrence after the preceding valid insertion. Thus, $\beta$ is an $R$-pattern.

Every root entry used by the occurrence lies to the left of $g$.  Let $p$
be the number of inserted entries it uses.  If $p\leq k$, the same root
entries together with $x_1,\ldots,x_p$ would already form $\beta$ after the
$p$th insertion.  Hence, $p=k+1$: all inserted entries are used and play the
$p$ largest entries of $\beta$.  They form an increasing suffix, while the
root entries form $\beta^{-p}$.  Since the first insertion was valid,
$p\geq2$.  Reflecting positions proves the decreasing-run statement.
\end{proof}

\begin{proof}[Proof of Theorem~\ref{thm:maximum-types}]
If there is no $R$-pattern, inserting every new maximum in the rightmost gap
can never create a forbidden occurrence.  Every node is therefore viable.
The case of no $L$-pattern is symmetric.

Suppose that $B$ has no $M$-pattern, and let $\rho$ be a pocket root.  If
$\rho$ has no active gap, its pocket has depth zero.  Otherwise, all increasing runs
from $\rho$ are finite, because $\rho$ is mortal.  Apply
Lemma~\ref{lem:one-sided-blocking} to the increasing run at the leftmost
active gap.  It gives an occurrence $T_R$ of $\beta_R^{-p}$ to the left of
every active gap, where the $p$ largest entries of $\beta_R$ form an
increasing suffix.  Applying the symmetric statement at the rightmost
active gap gives an occurrence $T_L$ of $\beta_L^{-q}$ to the right of
every active gap, where the $q$ largest entries form a decreasing prefix.
Here $2\leq p,q\leq \mm{B}$.

Take a longest path from $\rho$ to a leaf of its pocket, and read its newly
inserted entries from left to right as a sequence $X$.  
Every insertion lies between the fixed
root occurrences $T_R$ and $T_L$.  An increasing subsequence of length $p$
in $X$ would complete $T_R$ to an occurrence of $\beta_R$; a decreasing
subsequence of length $q$ would complete $T_L$ to an occurrence of
$\beta_L$.  Erd\H{o}s--Szekeres (Theorem~\ref{thm:erdos-szekeres}) therefore gives
\[
 |X|\leq(p-1)(q-1)\leq(\mm{B}-1)^2.
\]
This is the depth of the pocket.
\end{proof}

Together with Corollary~\ref{cor:bounded-pocket-depth-consequences}, these
three results (Theorem~\ref{thm:wqo_bounded_pocket_depth}, Theorem~\ref{thm:regular-bounded-pocket-depth}, and Theorem~\ref{thm:maximum-types}) prove the structural part of Theorem~\ref{thm:main-positive}.

\section{NP-completeness for a fixed basis}
\label{sec:hardness}

We now describe the fixed forbidden set
of permutations $\Bfinal$ and the reduction proving the NP-hardness of $\ExtendMax_{\Bfinal}^{\mathrm{un}}$
(Theorem~\ref{thm:main-hardness}).  The main text gives the construction
principles and the correctness argument.  Appendix~\ref{chap:extendmax}
provides all pattern definitions, coordinates, and occurrence proofs.

\subsection{Boolean states in the insert-maximum tree}
The source problem is monotone NAE-3SAT: every clause contains three
unnegated variable occurrences, and a satisfying assignment must give each
clause both a zero and a one.  This problem is
NP-complete~\cite[Theorem~2.1]{Schaefer1978}.  Let $\Phi$ be an instance of this problem with $n$ variables $x_1,\dots,x_n$ and $q$
clauses $C_1,\dots,C_q$, each of the form $(x_i,x_j,x_k)$ for $i,j,k \in \{1,\dots,n\}$.
We may assume that each of the variables appears in a clause. 
The empty formula is mapped to the Yes-instance
$(\varepsilon,1^0)$ of $\ExtendMax_{\Bfinal}^{\mathrm{un}}$, so we assume $q\ge1$.  If $n=1$,
then $\Phi$ is unsatisfiable.  Replacing it by the fixed unsatisfiable instance
$(x,x,x)\wedge(y,x,x)$ preserves the answer.  We therefore assume $n\ge2$.

A \emph{block} of a permutation is a set of
entries whose positions and values each form
a set of consecutive integers.
It is a \emph{$\tau$-block} if its entries
form an occurrence of $\tau$.
A permutation $\pi$ is \emph{simple} if every block of $\pi$ has size $0$, $1$, or $|\pi|$.
Let $\sigma=\sigma_1\cdots\sigma_m$ be a permutation.
The \emph{inflation}
$\sigma[\alpha_1,\ldots,\alpha_m]$
is the permutation obtained from $\sigma$ by
replacing each entry $\sigma_i$ by a block
inducing $\alpha_i$.
These blocks appear from left to right in
the order $i=1,\ldots,m$, and every entry
of the $i$-th block is smaller than every
entry of the $j$-th block exactly when
$\sigma_i<\sigma_j$~\cite{AlbertAtkinson2005}.
The direct sum $\oplus$ is defined as
$\alpha\oplus\beta=12[\alpha,\beta]$.

\begin{figure}[t]
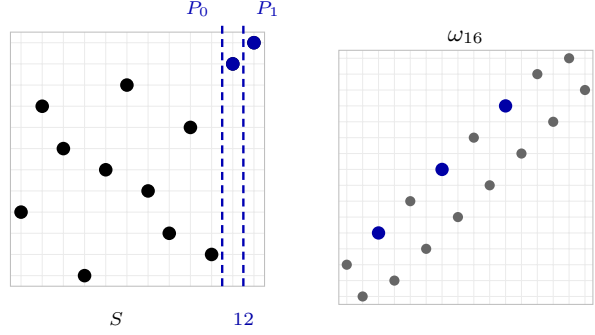

\centering
\begin{permpic}[0.28]{12}
  \pts{4,9,7,1,6,10,5,3,8,2,11,12}
  \foreach \x/\y in {11/11,12/12}
    {\fill[blue!65!black] (\x,\y) circle[radius=0.32];}
  \draw[blue!70!black,densely dashed,line width=0.9pt]
    (10.5,0.5)--(10.5,12.8);
  \draw[blue!70!black,densely dashed,line width=0.9pt]
    (11.5,0.5)--(11.5,12.8);
  \node[font=\scriptsize,blue!70!black,anchor=south east]
    at (10.35,12.8) {$P_0$};
  \node[font=\scriptsize,blue!70!black,anchor=south west]
    at (11.65,12.8) {$P_1$};
  \node[font=\scriptsize] at (5.5,-1.05) {$S$};
  \node[font=\scriptsize,blue!65!black] at (11.5,-1.05) {$12$};
\end{permpic}
\hspace{1.2em}
\begin{permpic}[0.21]{16}
  \pts[black!60]{3,1,5,2,7,4,9,6,11,8,13,10,15,12,16,14}
  \foreach \x/\y in {3/5,7/9,11/13}
    {\fill[blue!65!black] (\x,\y) circle[radius=0.42];}
  \node[font=\small] at (8.5,17.35) {$\omega_{16}$};
\end{permpic}
\caption{Left: the state permutation
$F=S\oplus12$. The black dots represent
the entries of $S$, the blue dots represent
the final increasing pair $12$, and the
dashed lines mark the insertion sites
$P_0$ and $P_1$.
Right: the increasing oscillation
$\omega_{16}$. The blue dots mark the
entries replaced by copies of $F$ in
$\rho_3^\star$.}
\label{fig:overview}
\end{figure}

We use the following four simple \emph{marker permutations} of length $10$:
\[
\begin{aligned}
D&=9,3,10,8,6,4,2,5,1,7,&
S&=4,9,7,1,6,10,5,3,8,2,\\
E&=5,9,3,8,10,7,1,6,4,2,&
N&=8,1,7,5,10,4,9,3,6,2.
\end{aligned}
\]
They are pairwise distinct and have longest decreasing subsequence length
$6$.  
Put $J:=12$ and define the \emph{state permutation} $F:=S\oplus J$
(see Figure~\ref{fig:overview}, left side).
The gaps $P_0 := 10$ and $P_1 := 11$ of $F$ are called the \emph{insertion sites}: a new maximum at $P_b$ represents
the bit $b$.

For the moment, let $r\geq1$ be arbitrary; it denotes the
number of state blocks that we want to place in one permutation.
To arrange $r$ such blocks, let $\omega_m$ be the \emph{(increasing) oscillation}
obtained by standardizing the first $m$ entries of $4,1,6,3,8,5,10,7,\ldots$ (see Figure~\ref{fig:overview}, right side). 
The \emph{base permutation} is the inflation
\[
\rho_r^\star
=\omega_{4r+4}[\alpha_1,\ldots,\alpha_{4r+4}],
\qquad
(\alpha_1,\ldots,\alpha_{4r+4})
=(D,N,(F,N,N,N)^r,E,N).
\]
It contains $r$ state blocks and $3r+4$ further marker blocks, so
$|\rho_r^\star|=42r+40$. For each $i\in[4r+4]$, let $I_i$ be the
block replacing the $i$-th entry of
$\omega_{4r+4}$ in this inflation.
We call $I_i$ the \emph{copy of $\alpha_i$}
corresponding to position $i$.
For $0\leq c<r$, let
$F(c):=I_{4c+3}$ be the $c$-th
\emph{state block}, and let $S(c)$ and
$J(c)$ consist of its first ten and
last two entries, respectively, in
positional order.
We call the sets $S(c)$, for $0\leq c<r$,
the \emph{copies of $S$}, and similarly
the sets $J(c)$ the \emph{copies of $J$}.
The copies of $D$, $N$, $E$, and $S$
are called the \emph{marker copies}.
Define
\[
\begin{aligned}
A&:=3142[F,N,N,N], & U&:=3142[N,N,F,N],\\
U_D&:=3142[D,N,F,N], & U_E&:=3142[N,N,E,N].
\end{aligned}
\]
For $0\leq c<r$, the four copies replacing
positions $4c+3,\ldots,4c+6$ of
$\omega_{4r+4}$ form the \emph{state window}
of $F(c)$.
For $0\leq c<r-1$, the four copies replacing
positions $4c+5,\ldots,4c+8$ form the
\emph{overlap window} between $F(c)$ and
$F(c+1)$.
The first and last four copies form the
\emph{left boundary window} and
\emph{right boundary window}, respectively.
These windows induce $A$, $U$, $U_D$,
and $U_E$, respectively.

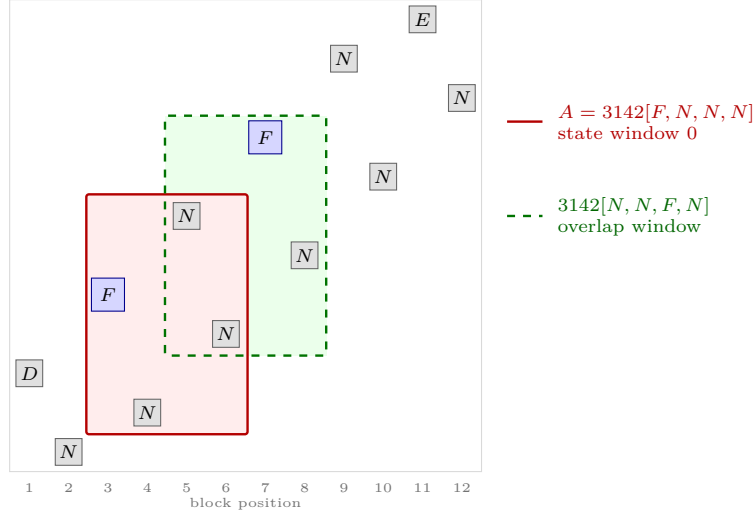
\begin{figure}[t]
\centering
\begin{tikzpicture}[x=0.52cm,y=0.52cm,every node/.style={font=\scriptsize}]
  \fill[green!8] (4.45,3.45) rectangle (8.55,9.55);
  \fill[red!7] (2.45,1.45) rectangle (6.55,7.55);
  \draw[gray!30] (0.5,0.5) rectangle (12.5,12.5);
  \foreach \i/\v/\lab in {1/3/D,2/1/N,4/2/N,5/7/N,6/4/N,
                            8/6/N,9/11/N,10/8/N,11/12/E,12/10/N}
    {\draw[fill=black!12,draw=black!60]
       (\i-0.34,\v-0.34) rectangle (\i+0.34,\v+0.34);
     \node at (\i,\v) {$\lab$};}
  \foreach \i/\v in {3/5,7/9}
    {\draw[fill=blue!16,draw=blue!55!black]
       (\i-0.42,\v-0.42) rectangle (\i+0.42,\v+0.42);
     \node at (\i,\v) {$F$};}
  \foreach \i in {1,...,12}
    {\node[font=\tiny,text=black!55] at (\i,0.08) {$\i$};}
  \node[font=\tiny,text=black!55] at (6.5,-0.32) {block position};
  \draw[red!70!black,line width=0.9pt,rounded corners=1pt]
     (2.45,1.45) rectangle (6.55,7.55);
  \draw[green!45!black,dashed,line width=0.9pt,rounded corners=1pt]
     (4.45,3.45) rectangle (8.55,9.55);
  \draw[red!70!black,line width=0.9pt] (13.15,9.4)--(14.05,9.4);
  \node[anchor=west,align=left,text=red!70!black] at (14.2,9.4)
     {$A=3142[F,N,N,N]$\\state window $0$};
  \draw[green!45!black,dashed,line width=0.9pt] (13.15,7.0)--(14.05,7.0);
  \node[anchor=west,align=left,text=green!40!black] at (14.2,7.0)
     {$3142[N,N,F,N]$\\overlap window};
\end{tikzpicture}
\caption{The base permutation $\rho^{\star}_2$: the plot of
$\omega_{12}$ with each entry replaced by its corresponding block.  Blue
boxes are state blocks and gray boxes are marker blocks.  The red box is
the first state window.  The green overlap window consists of the final two
$N$-copies of that window, the next state block, and its first following
$N$-copy.}
\label{fig:base-permutation}
\end{figure}

To allow at most one
insertion in each state block, exclusively at the insertion sites $P_0$ and $P_1$ of each state block, we forbid a fixed set of permutations $\Bstate$, which we will specify next. 
Let
$\ins^{\nearrow2}_g(\pi)$ be the permutation obtained from $\pi$ by inserting two increasing new maxima into the same original gap $g$. For $g_1<\cdots<g_t$, let
$\ins^\nearrow_{g_1,\ldots,g_t}(\pi)$
be obtained by inserting $t$ new maxima,
increasing from left to right, into the
indicated original gaps.
The set
$\Bstate$ is the union of the following families:
\[
\begin{aligned}
\mathcal B_{\mathrm{core}}
 &=\{\ins_g(T):T\in\{D,S,E,N\},\ 1\leq g\leq9\},\\
\mathcal B_{\mathrm{gap}}
 &=\{\ins_{12}(A),\ins_{22}(A)\},\\
\mathcal B_{\mathrm{cap}}
 &=\{\ins^{\nearrow2}_{10}(A),\ins^{\nearrow2}_{11}(A)\},\\
\mathcal B_{\mathrm{conf}}
 &=\{\ins^{\nearrow}_{10,11}(A)\},\\
\mathcal B_{\mathrm{overlap}}
 &=\{\ins_{10}(U),\ins_{20}(U)\},\\
\mathcal B_{\mathrm{boundary}}
 &=\{\ins_0(D),\ins_{10}(U_D),\ins_{20}(U_D),\\
 &\hspace{1.5em}\ins_{10}(U_E),\ins_{20}(U_E),
                    \ins_{30}(U_E),\ins_{40}(U_E)\},\\
\mathcal B_{\mathrm{order}}
 &=\{4312[1,1,N,E]\}.
\end{aligned}
\]
The \emph{core patterns} $\mathcal B_{\mathrm{core}}$ close the interior gaps of marker copies.  The \emph{gap and
overlap patterns} $\mathcal B_{\mathrm{gap}} \cup \mathcal B_{\mathrm{overlap}}$ close gaps $12,22,32,42$ in a state window. The \emph{boundary
patterns} $\mathcal B_{\mathrm{boundary}}$ deal with gaps at the ends.  Only sites $10$ and $11$ remain. The \emph{capacity patterns} $\mathcal B_{\mathrm{cap}}$ forbid two
insertions at the same site, whether that site is $10$ or $11$.
The \emph{conflict pattern} in $\mathcal B_{\mathrm{conf}}$ forbids one insertion
at site $10$ and another at site $11$ of the same state block.
The \emph{ordering pattern} in $\mathcal B_{\mathrm{order}}$ forces all inserted maxima to increase from left to right. 

\begin{lemma}[State behavior]
\label{lem:state-summary}
Let $r \geq 1$.
In $\Av(\Bstate)$, the active gaps of $\rho_r^\star$ are exactly the two sites of each
state block. Every $\Bstate$-avoiding extension inserts entries only at these
sites and contains at most one inserted entry in each state block. Moreover,
for each $(b_0,\ldots,b_{r-1})\in\{0,1\}^r$, inserting a maximum
at site $P_{b_c}$ of block $c$, processing the blocks from left to right,
defines a bijection between $\{0,1\}^r$ and the depth-$r$ extensions of
$\rho_r^\star$. In particular,
$d_{\Bstate}(\rho_r^\star)=r$.
\end{lemma}

\begin{proof}
The core, gap, overlap, and boundary patterns make every root gap
except the two sites of each state block inactive. Since an inactive root
gap remains inactive after further insertions, every inserted entry in an
avoiding extension must lie at a site.

The $N$-copy immediately before $E$ and the $E$-copy lie to the right
of all sites. Two
inserted entries in decreasing left-to-right order would therefore form
$4312[1,1,N,E]$ together with these copies. Hence, the inserted entries
increase from left to right. Lemmas~\ref{ext:lem:isolation},~\ref{ext:lem:window-occurrences}, and~\ref{ext:lem:boundary-occurrences} in Appendix~\ref{chap:extendmax} then show that every occurrence of a
marker permutation coincides with a marker copy.
Consequently, occurrences of $A$ and $U_E$ are designated windows, and
occurrences of $U$ and $U_D$ agree with designated windows on their
marker entries. The capacity patterns therefore forbid two entries at the same
site, while the conflict pattern forbids using both sites of one block.
Thus every extension has depth at most $r$.

Fix $(b_0,\ldots,b_{r-1})\in\{0,1\}^r$ and insert a
maximum at site $P_{b_c}$ of block $c$, processing the blocks from left to
right. The resulting permutation avoids $\Bstate$ and is therefore an
extension of depth $r$.

Conversely,  let $\sigma$ be any extension of $\rho_r^\star$ of depth $r$. Since every inserted
entry lies at a site and each state block contains at most one inserted entry,
$\sigma$ uses exactly one site in every state block. Write $P_{b_c}$ for the site
used in block $c$.
Then $\sigma$ is exactly the extension constructed from
$(b_0,\ldots,b_{r-1})$. Thus, the construction is a bijection. Full details
are given in Lemmas~\ref{ext:lem:ordering}--\ref{ext:lem:capacity}.
\end{proof}

\subsection{Routing and clause constraints}
\label{sect:clauses}

Let $x_{i_1},\dots,x_{i_h}$, where $h = 3q$, denote the appearances of the variables in $\Phi$
 in the order in which they appear in the clauses $C_1,\dots,C_q$.  
 Relabel the variables so that $i_h= n$.
 Put
$\lambda:=\lceil\log_2 n\rceil$ and $r:=(\lambda+2)h+n$. We partition the state blocks of
$\rho_r^\star$ into $\lambda+3$ \emph{levels}. 
The levels are convex in the sense that between the positions of blocks on the same level there are no positions of blocks from another level.
Levels
$0,\ldots,\lambda+1$ contain one state block for each variable occurrence, while level
$\lambda+2$ contains one state block for each variable. 
For $\ell \in \{0,\dots,\lambda+1\}$ and 
$u \in \{1,\dots,h\}$, 
let $F(\ell,u)$ denote the $u$-th state block on level $\ell$ (here we refer to the order of the state blocks in $\rho_r^\star$).
Levels \(0\) and \(1\) have their state blocks in clause order: for every \(\ell\in\{0,1\}\) and \(u\in\{1,\ldots,h\}\), the block \(F(\ell,u)\) represents the occurrence \(x_{i_u}\).

Give each variable \(x_j\) a \(\lambda\)-bit code: for \(1\leq j<n\), the code of \(x_j\) is the binary representation of \(j-1\), padded with leading zeros, while the code of \(x_n\) consists of \(\lambda\) ones. Number the bit positions from \(0\), starting with the least significant bit. To obtain level \(a+2\) from level \(a+1\), first list, in their order on level \(a+1\), all blocks representing occurrences of variables whose code has \(0\) in bit position \(a\); then list, again in their order on level \(a+1\), all blocks representing occurrences of variables whose code has \(1\) in bit position \(a\).
After the $\lambda$ passes,
all occurrences of each variable are consecutive. 

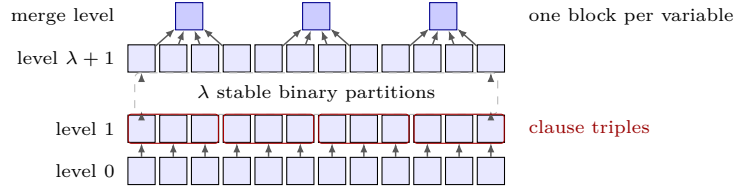
\begin{figure}[t]
\centering
\begin{tikzpicture}[x=0.42cm,y=0.62cm,every node/.style={font=\scriptsize},
   b/.style={draw,minimum width=3.6mm,minimum height=3.6mm,inner sep=0pt,fill=blue!9},
   active/.style={b,fill=blue!20,draw=blue!55!black},
   ar/.style={-{Latex[length=1.25mm]},black!65,line width=0.55pt}]
 \foreach \i in {0,...,11} {\node[b] (a\i) at (\i,0) {};}
 \foreach \i in {0,...,11} {\node[b] (b\i) at (\i,0.9) {};}
 \foreach \i in {0,...,11} {\node[b] (c\i) at (\i,2.4) {};}
 \node[active] (d0) at (1.5,3.3) {};
 \node[active] (d1) at (5.5,3.3) {};
 \node[active] (d2) at (9.5,3.3) {};
 \foreach \i in {0,...,11} {\draw[ar] (a\i)--(b\i);}
 \node[draw=gray!55,dashed,rounded corners=2pt,fill=white,
       minimum width=48mm,minimum height=5.5mm,align=center]
       at (5.5,1.65) {$\lambda$ stable binary partitions};
 \foreach \i in {0,11} {\draw[ar] (b\i)--(\i,1.28);
                           \draw[ar] (\i,2.02)--(c\i);}
 \foreach \s in {0,...,3} {\draw[ar] (c\s)--(d0);}
 \foreach \s in {4,...,7} {\draw[ar] (c\s)--(d1);}
 \foreach \s in {8,...,11} {\draw[ar] (c\s)--(d2);}
 \node[anchor=east] at (-0.55,0) {level $0$};
 \node[anchor=east] at (-0.55,0.9) {level $1$};
 \node[anchor=east] at (-0.55,2.4) {level $\lambda+1$};
 \node[anchor=east] at (-0.55,3.3) {merge level};
 \foreach \x in {-0.42,2.58,5.58,8.58}
   {\draw[red!65!black,rounded corners=1.2pt] (\x,0.59) rectangle ({\x+2.84},1.21);}
 \node[red!60!black,anchor=west] at (11.9,0.9) {clause triples};
 \node[anchor=west] at (11.9,3.3) {one block per variable};
\end{tikzpicture}
\caption{The routing for an example with $h=12$ and three variables.
Level $1$ contains the clause triples, the middle box represents the stable
binary partitions, and the final transition joins all occurrences of a
variable to its merge block.}
\label{fig:routing}
\end{figure}

A \emph{wire} $s\to t$ is an ordered pair of
state-block indices, requiring the bits of
$F(s)$ and $F(t)$ to agree.
We call $s$ its \emph{source} and $t$
its \emph{target}.
For $0\leq\ell\leq\lambda$, include a
wire between the blocks representing the
same variable occurrence on levels $\ell$
and $\ell+1$, directed towards the latter.
Also include a wire from each occurrence
block on level $\lambda+1$ to the block
representing its variable on level
$\lambda+2$.
For the formal details on how to implement wires, see Section~\ref{subsec:wires}.
At each reordering, the positions of the
blocks on the preceding level, read in
their new order, form two increasing
subsequences. Their longest decreasing
subsequence therefore has length at most two.
Figure~\ref{fig:routing} summarizes the routing.
The blocks on level $1$ are partitioned
into consecutive triples,
called \emph{clause triples}, whose assigned
bits are required to be not all equal. Section~\ref{subsec:wires} also explains how to implement these. 

\FloatBarrier

\subsection{Wires and their encoding}
\label{subsec:wires}

We now build $\rho_\Phi$ from $\rho_r^\star$ to implement the wires and clause
triples.  We use four further simple permutations of length $10$:
\[
\begin{aligned}
Q_L&=8,10,7,5,3,9,2,6,1,4,&
Q_R&=3,5,10,8,6,4,2,9,1,7,\\
C_L&=5,10,2,8,6,4,7,3,1,9,&
C_R&=8,10,7,4,6,1,5,3,9,2.
\end{aligned}
\]
They are pairwise distinct, distinct from the four marker permutations, and
have longest decreasing subsequence length $6$.

\begin{figure}[ht]
\centering
\begin{tikzpicture}[x=0.78cm,y=0.43cm,every node/.style={font=\scriptsize}]
  \draw[-{Latex[length=1.5mm]},black!45] (-0.55,-0.55)--(10.85,-0.55)
    node[anchor=west] {block position};
  \draw[-{Latex[length=1.5mm]},black!45] (-0.55,-0.55)--(-0.55,7.85)
    node[anchor=south,rotate=90] {block value};
  \draw[black!60] (0,0) rectangle (3.20,3.20);
  \node[anchor=south west,font=\tiny] at (0.05,3.26) {source $F(s)$};
  \draw[fill=black!10,draw=black!55] (0.18,0.18) rectangle (1.52,0.95);
  \node at (0.85,0.565) {$S(s)$};
  \draw[dashed,orange!70!black,fill=orange!12]
    (0.18,1.15) rectangle (3.02,2.30);
  \node[orange!60!black,align=center,font=\tiny] at (1.60,1.725)
    {values between\\$S(s)$ and $J(s)$};
  \draw[fill=blue!12,draw=blue!55!black] (1.72,2.48) rectangle (3.02,3.02);
  \node at (2.37,2.75) {$J(s)$};
  \draw[black!60] (5.00,4.30) rectangle (8.20,7.50);
  \node[anchor=south west,font=\tiny] at (5.05,7.56) {target $F(t)$};
  \draw[fill=black!10,draw=black!55] (5.18,4.48) rectangle (6.52,5.25);
  \node at (5.85,4.865) {$S(t)$};
  \draw[fill=blue!12,draw=blue!55!black] (6.72,6.58) rectangle (8.02,7.12);
  \node at (7.37,6.85) {$J(t)$};
  \draw[fill=orange!28,draw=orange!70!black]
    (4.00,1.30) rectangle (4.85,1.88);
  \node[font=\tiny] at (4.425,1.59) {$Q_L(s)$};
  \draw[fill=orange!28,draw=orange!70!black]
    (8.35,1.70) rectangle (9.20,2.28);
  \node[font=\tiny] at (8.775,1.99) {$Q_R(s)$};
  \draw[dotted,orange!75!black,line width=0.8pt]
    (3.02,1.59)--(4.00,1.59);
  \draw[dotted,orange!75!black,line width=0.8pt]
    (3.02,1.99)--(8.35,1.99);
  \draw[black!45,dashed,-{Latex[length=1.3mm]}]
    (4.425,2.02)--(4.425,3.55)--(5.00,4.20);
  \draw[black!45,dashed,-{Latex[length=1.3mm]}]
    (8.775,2.42)--(8.775,3.55)--(8.20,4.20);
  \node[fill=white,inner sep=1.5pt] at (6.60,3.50)
    {positionally bracket $F(t)$};
  \draw[<->,black!55,line width=0.7pt] (9.75,2.30)--(9.75,4.30);
  \node[anchor=west,align=left] at (9.92,3.30)
    {source window lies below\\the target window};
\end{tikzpicture}
\caption{The copies $Q_L(s)$ and $Q_R(s)$ lie between $S(s)$ and
$J(s)$ in value, and 
$S(t)$ and $J(t)$ lie between 
$Q_L(s)$ and $Q_R(s)$ 
 in position.  The source window lies
below the target window.}
\label{fig:wire}
\end{figure}
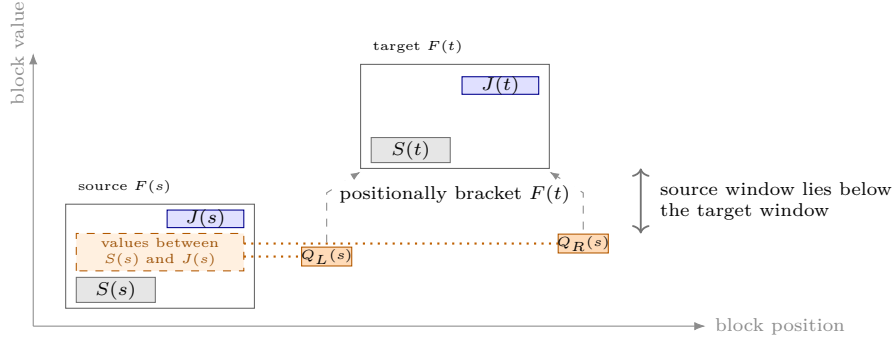

A wire $s\to t$ is encoded by placing copies of $Q_L$ and $Q_R$ around
$F(t)$ in position and between $S(s)$ and $J(s)$ in value (see Figure~\ref{fig:wire}). 
The value
placement identifies the source $s$, while the positional placement
identifies the target $t$.  Similarly, copies of $C_L$ and $C_R$ are placed
on opposite sides of the three blocks of a clause triple.  For sets of
entries $U$ and $V$, write $U<V$ when every value in $U$ is smaller than
every value in $V$.
Every source block has exactly one outgoing wire, so the two copies encoding
a wire can be indexed by its source. 

\begin{definition}\label{def:rhoPhi}
Define the permutation $\rho_\Phi$
from $\rho_r^\star$ by the following
three rules. The copies of $Q_L$,
$Q_R$, $C_L$, and $C_R$ introduced
below are called the \emph{constraint
copies}.
\begin{enumerate}[label=(D\arabic*), ref=D\arabic*]
\item
For each state block $t$ with incoming
wires $s_1\to t,\ldots,s_m\to t$,
list the sources in their left-to-right
order.
For every $i$, introduce two disjoint
sets of ten new entries, denoted by
$Q_L(s_i)$ and $Q_R(s_i)$, inducing
$Q_L$ and $Q_R$, respectively.
Place
$Q_L(s_1),\ldots,Q_L(s_m)$
consecutively immediately before $F(t)$
and
$Q_R(s_1),\ldots,Q_R(s_m)$
consecutively immediately after $F(t)$.
Place their values so that
\[
S(s_i)<Q_L(s_i)<Q_R(s_i)<J(s_i).
\]
\item
For each clause triple $(t_0,t_1,t_2)$,
let $s_i\to t_i$ be the incoming wire
for $i\in\{0,1,2\}$.
Introduce two disjoint sets of ten new
entries, denoted by $C_L(s_0)$ and
$C_R(s_0)$, inducing $C_L$ and $C_R$,
respectively.
Place $C_L(s_0)$ immediately before
$Q_L(s_0)$ and $C_R(s_0)$ immediately
after $Q_R(s_2)$.
Place their values so that
\[
S(s_0)<C_L(s_0)<Q_L(s_0)
<Q_R(s_0)<C_R(s_0)<J(s_0).
\]
\item
Add no other constraint copies.
\end{enumerate}

All constraint copies are pairwise
disjoint and occupy consecutive positions
and consecutive values in $\rho_\Phi$.
These rules preserve the relative position
and value orders of the original
$\rho_r^\star$ entries and determine
$\rho_\Phi$ uniquely.
\end{definition}
Figure~\ref{fig:wire} shows the geometry of one wire.

\FloatBarrier

There are $(\lambda+2)h$ wires and $q$ clause triples.  Consequently,
$|\rho_\Phi|=(42r+40)+20(\lambda+2)h+20q=O(h\log h)$.
The permutation $\rho_\Phi$ and the unary target $1^r$ can therefore be constructed in
polynomial time.

\subsection{Four violation patterns}

For a wire $s\to t$, take the four blocks of the source window, the copy
$Q_L(s)$, the state block $F(t)$, the copy $Q_R(s)$, and the three
$N$-copies of the target window, in positional order.  These $104$ entries
induce a fixed permutation $\Aedge$, independent of the chosen wire.  The
source sites are gaps $10,11$ and the target sites are gaps $62,63$ in
$\Aedge$.  Define
\[
 \Pneq{01}=\ins^{\nearrow}_{10,63}(\Aedge),
 \qquad
 \Pneq{10}=\ins^{\nearrow}_{11,62}(\Aedge).
\]
Both patterns have
length $106$ and represent the two unequal assignments to the endpoints.

For a clause triple $(t_0,t_1,t_2)$, let $s_0$ be the source of the incoming
wire at $t_0$.  Take the source window, the copy $C_L(s_0)$, the complete
windows of $t_0$ and $t_1$, the state block $F(t_2)$, the copy $C_R(s_0)$,
and the three $N$-copies of the window of $t_2$, in positional order.  These
$188$ entries induce a fixed permutation $\Anae$, independent of the chosen
clause triple.  The sites of the three target blocks are gaps
$62,63$, $104,105$, and $146,147$.  Define
\[
 \Pmono{000}=\ins^{\nearrow}_{62,104,146}(\Anae),
 \qquad
 \Pmono{111}=\ins^{\nearrow}_{63,105,147}(\Anae).
\]
Both patterns
have length $191$ and represent the two constant assignments to a clause
triple. 
The independence of the standardizations from the chosen wire or clause
is proved in Lemmas~\ref{ext:lem:Aeq-welldef}
and~\ref{ext:lem:Anae-welldef}.
Finally, define our fixed forbidden set 
\[
 \Bfinal:=\Bstate\cup
 \{\Pneq{01},\Pneq{10},\Pmono{000},\Pmono{111}\}.
\]

\subsection{No unintended occurrences}

The following lemma summarizes the properties of $\rho_\Phi$ needed
for correctness. For each $(b_0,\ldots,b_{r-1})\in\{0,1\}^r$,
start with $\rho_\Phi$ and insert one maximum
at site $P_{b_c}$ of each state block $F(c)$,
processing the sites from left to right.
We call the resulting permutation an
\emph{assignment configuration} of $\rho_\Phi$. 

\begin{lemma}[Properties of $\rho_\Phi$]
\label{lem:recognition-summary}
The permutation $\rho_\Phi$ (Definition~\ref{def:rhoPhi}) has the following properties.
\begin{enumerate}[label=(\roman*)]
\item $\rho_\Phi\in\Av(\Bfinal)$ and
      $d_{\Bstate}(\rho_\Phi)=r$.
\item The depth-$r$ $\Bstate$-avoiding extensions of $\rho_\Phi$ are
      exactly the assignment configurations of $\rho_\Phi$.
\item Let $\pi$ be an assignment configuration of $\rho_\Phi$.  For each
      $(a,b)\in\{(0,1),(1,0)\}$, the pattern $\Pneq{ab}$ occurs in $\pi$
      exactly when some wire $s\to t$ satisfies $(b_s,b_t)=(a,b)$.
      For each $b\in\{0,1\}$, the pattern $\Pmono{bbb}$ occurs in $\pi$
      exactly when some clause triple $(t_0,t_1,t_2)$ satisfies
      $(b_{t_0},b_{t_1},b_{t_2})=(b,b,b)$.
\end{enumerate}
\end{lemma}

\begin{proof}[Proof overview]
First let $\sigma$ be any $\Bstate$-avoiding extension of
$\rho_\Phi$. Deleting the constraint-copy entries in $\sigma$ and standardizing gives a
$\Bstate$-avoiding extension of $\rho_r^\star$ at the same depth.  Lemma~\ref{lem:state-summary}
therefore bounds the depth by $r$.  If the depth is $r$, that lemma also
shows that exactly one site $P_{b_c}$ is used in every state block, in
left-to-right order.  Hence, $\sigma$ is the corresponding assignment
configuration.
It remains to show that every assignment configuration $\pi$ avoids
$\Bstate$.  After deleting the constraint-copy entries and standardizing, we obtain the
depth-$r$ extension of $\rho_r^\star$ encoded by the same bits, which avoids
$\Bstate$ by Lemma~\ref{lem:state-summary}.  Thus, a new occurrence of a
member of $\Bstate$ would have to use an entry of a constraint copy.
The same occurrence results used in
Lemma~\ref{lem:state-summary} apply to $\pi$.
By Lemmas~\ref{ext:lem:isolation}, \ref{ext:lem:window-occurrences},
and~\ref{ext:lem:boundary-occurrences}, marker and constraint occurrences
are copies, occurrences of $A,U_E$ are windows, and occurrences of
$U,U_D$ match windows on marker entries.

Constraint-copy entries cannot create occurrences of members of $\Bstate$.  Constraint copies lie neither inside marker copies nor at
sites, excluding the core, capacity, and conflict patterns.  In every
relevant gap of a state or overlap window, a constraint-copy entry lies
below the window, while the additional entry of the forbidden pattern lies
above it.  All constraint copies lie below the final copy of $E$, so the two entries
above $E$ in the ordering pattern would have to be inserted maxima, which
increase from left to right.  Finally, the placement rules leave the
relevant boundary gaps free of constraint copies.  Hence, $\pi$ avoids
$\Bstate$.  Such a configuration has depth $r$, while the first paragraph
gave the matching upper bound and characterized every extension at that
depth.  This proves $d_{\Bstate}(\rho_\Phi)=r$ and~(ii).  Deleting the
inserted maxima also shows that $\rho_\Phi$ avoids $\Bstate$.

Fix an assignment configuration $\pi$.  Remove the two largest
entries from an occurrence of $\Pneq{ab}$ in $\pi$. The marker
and constraint occurrences within this
occurrence of $\Aedge$ are copies.  The $Q_L$- and $Q_R$-copies lie between $S(s)$ and $J(s)$ in value and
positionally bracket the target state block.  They therefore encode an
actual wire $s\to t$, and the removed entries are the inserted maxima at site $P_a$ of $F(s)$ and
site $P_b$ of $F(t)$.  Removing the three largest entries from an occurrence
of $\Pmono{bbb}$ similarly identifies an actual clause triple and shows that
the removed entries are its three inserted maxima at the $P_b$ sites.  Thus
every violation-pattern occurrence in $\pi$ uses inserted maxima.  If
$\rho_\Phi$ contained such a pattern, the same root entries would form an
occurrence in $\pi$, which is impossible.  Hence, $\rho_\Phi$ avoids the four
violation patterns.  Together with its $\Bstate$-avoidance, this proves
$\rho_\Phi\in\Av(\Bfinal)$ and completes~(i).

Conversely, every wire $s\to t$ with
bit $a$ at $s$ and bit $b$ at $t$,
where $a\ne b$, gives an occurrence
of $\Pneq{ab}$, and every clause triple
whose three blocks have bit $b$ gives
an occurrence of $\Pmono{bbb}$.  Combined
with the preceding recognition of all occurrences, this proves~(iii).
The complete arguments are given in
Theorem~\ref{ext:thm:preservation}, Corollary~\ref{ext:cor:state-depth},
Corollaries~\ref{ext:cor:edge-avoidance} and~\ref{ext:cor:nae-avoidance},
and Lemma~\ref{ext:lem:root-avoidance}.
\end{proof}

\subsection{Correctness and consequences}
Lemma~\ref{lem:recognition-summary}(ii)--(iii) shows that a
$\Bfinal$-avoiding extension of depth $r$ exists exactly when one can choose
one bit per state block, with equal bits on every wire and a nonconstant
triple in every clause. Such choices
exist exactly when $\Phi$ is NAE-satisfiable.

\begin{theorem}
\label{thm:formula-main}
We have $d_{\Bfinal}(\rho_\Phi)\leq r$, with equality if
and only if $\Phi$ is NAE-satisfiable.
\end{theorem}
\begin{proof}
Lemma~\ref{lem:recognition-summary}(i)
gives root avoidance and the depth
upper bound, as $\Bstate\subseteq\Bfinal$.
If $\Phi$ is satisfiable, choose a satisfying assignment and assign each
state block its variable's bit.
The bits agree along wires and are
nonconstant on clause triples, so
parts~(ii) and (iii) give a
$\Bfinal$-avoiding extension of depth $r$.

Conversely, by part~(ii), any avoiding
extension at depth $r$ uses exactly
one site per state block.
Part~(iii) forces bit agreement along
wires and nonconstant clause triples.
The wire paths to merge blocks then
force all occurrences of each variable
to agree, yielding a NAE-satisfying
assignment of $\Phi$.
\end{proof}

\begin{proof}[Proof of Theorem~\ref{thm:main-hardness}]
Map $\Phi$ to $(\rho_\Phi,1^r)$.  The size bounds make this a
polynomial-time reduction, and Theorem~\ref{thm:formula-main} proves its
correctness. Hence $\ExtendMax_{\Bfinal}^{\mathrm{un}}$ is NP-hard.
Replacing the unary representation of $r$ by its binary representation
gives NP-hardness of $\ExtendMax_{\Bfinal}^{\mathrm{bin}}$. Theorem~\ref{thm:extendmax-in-np} gives NP membership for
both encodings.
\end{proof}

By Proposition~\ref{prop:existence-enumeration-equivalence},
$\EnumExt_{\Bfinal}\notin\mathrm{DelayP}$ unless $\mathrm P=\mathrm{NP}$.
The class also has unbounded pocket depth, without any complexity
assumption.  Take satisfiable formulas with arbitrarily many disjoint
clauses.  Their roots are mortal and have unbounded finite depth $r$.
Each lies in a pocket of at least that depth.

Finally, consider Hanika's problem $\FPCSP(B)$.  Its input is a finite set
$V$ and two acyclic relations $E_1,E_2$ on $V$.  It asks whether linear
orders extending the relations induce a permutation in $\Av(B)$: label
the elements by their order in the first linear order and read them in the
second.
The following answers~\cite[Problem~4.1]{Hanika2015}.

\begin{corollary}
\label{cor:fpcsp-main}
$\FPCSP(\Bfinal)$ is NP-complete, even when the first order is fixed.
\end{corollary}
\begin{proof}
Given $(\rho,1^k)$ with $|\rho|=n$, set $V=\{1,\ldots,n+k\}$.
Let $E_1$ be the chain $1<\cdots<n+k$, and let $E_2$ require
$\rho_1<\cdots<\rho_n$, with the remaining elements unconstrained.
The induced permutations are exactly the insert-maximum extensions of
$\rho$ at depth $k$.  This is a polynomial-time reduction, because $k$ is represented in unary.  Membership in NP follows by guessing the two orders
and checking the relations and avoidance for fixed $B$.
\end{proof}

\section{Conclusion and Open Problems}
For every fixed finite set $B$ of forbidden permutations the problem $\EnumAvoid_B$ belongs to $\operatorname{OutputP}$. 
For some $B$, we even obtain a polynomial-delay algorithm. To prove such polynomial-delay results, we introduced the problem 
$\ExtendMax_B^{\mathrm{un}}$. 
If $\ExtendMax_B^{\mathrm{un}}$ is in P, then $\EnumAvoid_B$ is in $\mathrm{DelayP}$. 
We introduce and study \emph{pockets} and use this concept 
to formulate three structural conditions 
on $B$ that imply that $\ExtendMax_B^{\mathrm{un}}$ is in P. 
Finally, we find a finite set $\Bfinal$
of permutations such that
$\ExtendMax_{\Bfinal}^{\mathrm{un}}$
is NP-complete.  

The main remaining question is whether every finitely based permutation
class admits polynomial-delay avoider enumeration.  In particular, the
complexity of $\EnumAvoid_{\Bfinal}$ remains unresolved here.  Another
question is whether fixed-basis insert-maximum extension admits a
dichotomy between polynomial time and NP-completeness.

\appendix

Write $\LDS(\pi)$ for the lengths of the longest
increasing and decreasing subsequences of $\pi$.

\section{Full construction and proof of fixed-basis hardness}
\label{chap:extendmax}

In this appendix we provide the full details of the proof that
$\ExtendMax_{\Bfinal}$ is \NPhard{}.
As mentioned in the main text, 
the reduction is from monotone NAE-3SAT and builds the starting permutation
from Boolean state blocks arranged in levels.  The first level encodes the
clauses, successive levels reorder the variable occurrences, and a final
merge brings together the variable occurrences belonging to each variable.
Fixed forbidden patterns force every maximum-depth extension to choose one bit
in each state block, require all occurrences of each variable to choose the same bit, and enforce
the not-all-equal condition in every clause.  Thus, an extension of the target
depth exists exactly when the formula has a not-all-equal satisfying
assignment.

\subsection{Permutation operations used in the reduction}
\label{ext:sec:permutation-operations}
\begin{figure}[ht]
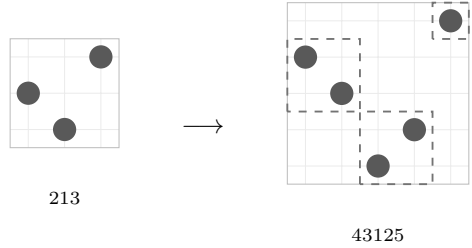

\centering
\begin{permpic}[0.48]{3}
  \pts[black!65]{2,1,3}
  \node[font=\scriptsize] at (2,-0.9) {$213$};
\end{permpic}
\qquad
$\longrightarrow$
\qquad
\begin{permpic}[0.48]{5}
  \pts[black!65]{4,3,1,2,5}
  \draw[black!55,dashed,line width=0.7pt] (0.5,2.5) rectangle (2.5,4.5);
  \draw[black!55,dashed,line width=0.7pt] (2.5,0.5) rectangle (4.5,2.5);
  \draw[black!55,dashed,line width=0.7pt] (4.5,4.5) rectangle (5.5,5.5);
  \node[font=\scriptsize] at (3,-0.9) {$43125$};
\end{permpic}
\caption{The inflation $213[21,12,1]=43125$.  The dashed rectangles mark
the three blocks, which induce $21$, $12$, and $1$, respectively.}
\label{ext:fig:inflation-example}
\end{figure}

The construction uses blocks, inflations, sums, and decreasing
subsequences. 
Blocks, simple permutations, and inflations were defined in
Section~\ref{sec:hardness}. For example, the entries $3$ and $2$ of
$1324$ form a block, while $2413$ is simple.
An example of an inflation is 
\[
  213[21,12,1]=43125.
\]
The three inflated blocks are shown in
Figure~\ref{ext:fig:inflation-example}.

\subsection{The source problem and reduction outline}
\label{ext:sec:reduction}

The reduction starts from monotone not-all-equal satisfiability.

\begin{definition}[Monotone NAE-3SAT]
\label{ext:def:source}
An instance is a formula $\Phi$ in conjunctive normal form (CNF) in which
every clause consists of exactly three unnegated variable occurrences.  The
question is whether there is a truth assignment that makes every clause
\emph{not-all-equal}; that is, each clause must contain at least one true and
at least one false variable occurrence.
\end{definition}

\begin{theorem}[Schaefer]
\label{ext:thm:source-hard}
The problem in Definition~\ref{ext:def:source} is
\NPcomplete{}~\cite[Theorem~2.1]{Schaefer1978}.
\end{theorem}

Let $\Phi$ have $n$ variables and $q$ clauses, and let $h$ be its number of
variable occurrences.  Then $h=3q$.
We take the variables of $\Phi$ to be exactly those that occur in the formula,
so $n\le h$.  The empty formula is mapped to the Yes-instance
$(\varepsilon,1^0)$, so we assume $q\ge1$ and hence $h\ge3$.  If $n=1$,
then $\Phi$ is unsatisfiable.  Replacing it by the fixed unsatisfiable instance
$(x,x,x)\wedge(y,x,x)$ preserves the answer.  We therefore assume $n\ge2$.

We now construct the permutation instance from a formula.  Three Boolean
requirements must be enforced.  Each variable occurrence must carry a bit.
The three variable occurrences belonging to a clause must not all carry the
same bit.  All variable occurrences belonging to any one variable must carry
a common bit.  The following sections address these requirements.

\subsection{Encoding a bit in a state block}
\label{ext:sec:construction}

We first construct a state block that carries one bit.  It contains two
possible insertion sites.  The forbidden patterns introduced below permit at most one
of them to be used, and the two choices encode $0$ and $1$.

\subsubsection{Marker permutations}
\label{ext:sec:markers}

\begin{definition}[Marker permutations]
\label{ext:def:markers}
Let $D,S,E,N\in\mathfrak S_{10}$ 
be the four permutations displayed and drawn in
Appendix~\ref{ext:app:fixed-permutation-data}.  We call them the \emph{marker permutations}.
\end{definition}

\begin{proposition}[Properties of the marker permutations]
\label{ext:prop:marker-facts}
The four marker permutations are pairwise distinct and simple, and every marker
permutation $T$ satisfies $\LDS(T)=6$.
\end{proposition}

\begin{proof}
A short program exhaustively checks the four displayed permutations of length
$10$ and confirms the stated properties.
\end{proof}

The copies of these permutations placed by the construction in the root are the
\emph{marker copies}.  The simplicity and common longest-decreasing-subsequence
length of the marker permutations are used later to identify every placed copy
uniquely.

\subsubsection{State blocks and their Boolean states}
\label{ext:sec:state-block}

Each state block consists of one copy of $S$ followed by a short
increasing tail.  Two gaps are potential insertion sites, and exactly one is
used in a maximum-depth extension.

\begin{definition}[State permutation and sites]
\label{ext:def:state-block}
Put $J=12\in\mathfrak S_2$ and define the \emph{state permutation}
\[
  F=S\oplus J=12[S,J]\in\mathfrak S_{12}.
\]
The two sites are
\[
  P_0=\text{gap }10,
  \qquad
  P_1=\text{gap }11.
\]
A maximum inserted at $P_0$ encodes $0$, and a maximum inserted at $P_1$
encodes $1$.  A site is \emph{used} if an inserted maximum is placed there.
A state block is assigned bit $b\in\{0,1\}$ when its set of used sites is
exactly $\{P_b\}$.
$F(s)$ denotes the state block indexed by $s$, and
$S(s)$ and $J(s)$ denote its designated copies of $S$ and $J$.
\end{definition}

Figure~\ref{fig:overview} (left) illustrates $F$ and its two sites.

\subsection{Building the base permutation}
\label{ext:sec:base-permutation}

To place many state blocks in one permutation, we arrange them along an
increasing oscillation.  This produces a base permutation with $r$
state blocks.  The construction will later determine $r$ from the formula and
assign a role to each state block.

\subsubsection{The increasing oscillation}
\label{ext:sec:oscillation-def}

Increasing oscillations arrange the state blocks in a repeated local pattern.
The four-entry pattern $\vartheta$ will be used to identify them.

\begin{definition}[Increasing oscillations]
\label{ext:def:oscillation}
Let
\[
a_1,a_2,a_3,\ldots=4,1,6,3,8,5,10,7,\ldots,
\]
where, for $k\ge 1$,
\[
a_{2k-1}=2k+2
\qquad\text{and}\qquad
a_{2k}=2k-1.
\]
For $m\ge 2$, put $\omega_m=\st(a_1,\ldots,a_m)$. 
Finally, put $\vartheta=3142=\omega_4$.
We call $(a_i)_{i\ge1}$ the \emph{defining sequence} of the oscillations.
\end{definition}
Figure~\ref{fig:overview} (right) illustrates an increasing oscillation.

We now inflate an increasing oscillation with state blocks and marker copies
to form the base permutation.

\begin{definition}[Base permutation]
\label{ext:def:base-permutation}
For $r\ge1$, define the \emph{base permutation}
\[
  \rho^{\star}_r
  =\omega_{4r+4}[D,N,(F,N,N,N)^r,E,N].
\]
The $r$ copies of $F$ are the \emph{state blocks}.  We index them by
$c=0,1,\dots,r-1$, so state block $c$ sits at position $4c+3$ of
$\omega_{4r+4}$.
\end{definition}

There are $r$ state blocks of length $12$ and $3r+4$ additional marker copies
of length $10$, so
\begin{equation}
\label{ext:eq:root-length}
  |\rho^{\star}_r|=42r+40.
\end{equation}

\subsubsection{State windows and their forbidden patterns}

The same forbidden patterns should control every state block independently of
its position in the base permutation.  We therefore use the state block together with
the three $N$-copies that follow it.

\begin{definition}[State window]
\label{ext:def:windows}
Set
\[
  A=\vartheta[F,N,N,N].
\]
Thus $A$ has length $12+3\cdot10=42$.  In $\rho^{\star}_r$, the \emph{state window}
of state block $c$ consists of the four copies at positions
$4c+3,\dots,4c+6$ of $\omega_{4r+4}$ and induces $A$.
\end{definition}

For state block $c$, exactly $20+42c$ entries precede its copy of $F$, so
its complete window occupies positions
\[
  20+42c+1,\dots,20+42c+42.
\]
Some of the forbidden patterns are obtained from a state window by adding two
or three new maxima.  In addition to the single-insertion notation
$\ins_g(\pi)$ from Section~\ref{sec:insert-maximum-tree}, we use the following
notation.  All gaps refer to the \emph{original} permutation $\pi$.
\begin{itemize}
\item $\ins^{\nearrow2}_g(\pi)$ inserts two new maxima into the same gap $g$,
      in increasing order from left to right.
\item For $g_1<\cdots<g_t$, the notation
      $\ins^{\nearrow}_{g_1,\ldots,g_t}(\pi)$ from the main text inserts
      $t$ new maxima into the specified gaps, increasing from left to right.
\end{itemize}

\smallskip

It is enough to prove that each unwanted root gap is inactive.
Indeed, an insertion into an inactive root gap in any extension would,
by deleting the other inserted entries and standardizing, yield a forbidden
one-step extension of the root.

The first family of forbidden patterns forbids insertions into the nine
interior gaps of the marker copies:
\[
  \mathcal B_{\mathrm{core}}
  =\{\ins_g(T):T\in\{D,S,E,N\},\ 1\le g\le9\}.
\]
Within $A$, gap $12$ follows $F$ and gap $22$ follows the first $N$-copy.
The next family makes precisely those two gaps inactive:
\[
  \mathcal B_{\mathrm{gap}}
  =\{\ins_{12}(A),\ins_{22}(A)\}.
\]
The capacity patterns
\[
  \mathcal B_{\mathrm{cap}}
  =\{\ins^{\nearrow2}_{10}(A),\ins^{\nearrow2}_{11}(A)\}
\]
prevent a site from being used twice, while the conflict pattern
\[
  \mathcal B_{\mathrm{conf}}
  =\{\ins^{\nearrow}_{10,11}(A)\}
\]
prevents both sites of one state block from being used.  These patterns are
written with increasing inserted maxima.  The ordering pattern defined below
will force that order.

\subsubsection{The remaining gaps}

The previous forbidden patterns leave the final two gaps of each state window
and the gaps at the two ends of the root.  The following patterns describe
the overlap and boundary windows that locate those gaps.

\begin{definition}[Overlap and boundary windows]
\label{ext:def:boundary-patterns}
Set
\[
  U=\vartheta[N,N,F,N],
  \qquad
  U_D=\vartheta[D,N,F,N],
  \qquad
  U_E=\vartheta[N,N,E,N].
\]
The patterns $U$ and $U_D$ have length $42$, while $U_E$ has length $40$.
The four designated copies at positions $4c+5,\ldots,4c+8$
form the overlap window between $F(c)$ and $F(c+1)$.
The first and last four designated copies form the left and right boundary windows, respectively. They induce $U$, $U_D$, and $U_E$ as appropriate.
\end{definition}

For two consecutive state windows, the occurrence of $U$ uses the final two
$N$-copies of the first window, the next state block, and its first following
$N$-copy.  Hence
\[
  \mathcal B_{\mathrm{overlap}}
  =\{\ins_{10}(U),\ins_{20}(U)\}
\]
makes gaps $32$ and $42$ of every nonfinal state window inactive.

The pattern $\ins_0(D)$ makes the initial gap of the root inactive.
The designated occurrence of $U_D$ in the base permutation consists of the
initial copies of $D,N$, the first state block, and its first following
$N$-copy.  Thus $\ins_{10}(U_D)$ makes the gap after $D$ inactive, and
$\ins_{20}(U_D)$ makes the gap between the initial $N$ and the first state
block inactive.  The designated occurrence of $U_E$ consists of the final
two $N$-copies of the last state window followed by $E,N$.  Its gaps $10$
and $20$ are gaps $32$ and $42$ of the final state window, gap $30$ is the
gap after $E$, and gap $40$ is the final gap of the root.  We therefore set
\[
\begin{aligned}
  \mathcal B_{\mathrm{boundary}}=\{&\ins_0(D),
  \ins_{10}(U_D),\ins_{20}(U_D),\\
  &\ins_{10}(U_E),\ins_{20}(U_E),
  \ins_{30}(U_E),\ins_{40}(U_E)\}.
\end{aligned}
\]

The third $N$-copy following the final state block and the final copy of $E$
also support the ordering pattern in
\[
  \mathcal B_{\mathrm{order}}=\{4312[1,1,N,E]\},
\]
which will force all inserted maxima to increase from left to right.

\begin{figure}[htbp]
\centering
\begin{tikzpicture}[x=0.92cm,y=0.92cm,every node/.style={font=\scriptsize}]
  \draw[red!70!black,line width=1.1pt] (0,2.05)--(0.65,2.05);
  \node[anchor=west,red!65!black] at (0.78,2.05)
    {made inactive by an $A$-based pattern};
  \draw[green!45!black,dashed,line width=1.1pt] (0,1.55)--(0.65,1.55);
  \node[anchor=west,green!35!black] at (0.78,1.55)
    {made inactive by a $U$- or $U_E$-based pattern};

  \draw[fill=black!10,draw=black!55] (0,0) rectangle (3.0,0.62);
  \node at (1.5,0.31) {$S$};
  \draw[fill=blue!12,draw=blue!50!black] (3.25,0) rectangle (3.70,0.62);
  \draw[fill=blue!12,draw=blue!50!black] (3.70,0) rectangle (4.15,0.62);
  \node[font=\tiny] at (3.475,0.31) {$1$};
  \node[font=\tiny] at (3.925,0.31) {$2$};
  \node[blue!55!black,anchor=south] at (3.70,0.72) {$J$};
  \foreach \a/\b in {4.50/6.80,7.10/9.40,9.70/12.00}
    {\draw[fill=black!10,draw=black!55] (\a,0) rectangle (\b,0.62);
     \node at ({(\a+\b)/2},0.31) {$N$};}

  \draw[blue!70!black,line width=1.0pt] (3.12,-0.12)--(3.12,0.75);
  \draw[blue!70!black,line width=1.0pt] (3.70,-0.12)--(3.70,0.75);
  \node[blue!60!black,anchor=north] at (3.16,-0.16) {$P_0$};
  \node[blue!60!black,anchor=north] at (3.70,-0.16) {$P_1$};

  \foreach \x/\lab in {4.32/{12},6.95/{22}}
    {\draw[red!70!black,line width=1.2pt] (\x,-0.12)--(\x,0.82);
     \node[red!65!black,anchor=south] at (\x,0.86) {$\lab$};}
  \foreach \x/\lab in {9.55/{32},12.15/{42}}
    {\draw[green!45!black,dashed,line width=1.2pt] (\x,-0.12)--(\x,0.82);
     \node[green!35!black,anchor=south] at (\x,0.86) {$\lab$};}

  \draw[decorate,decoration={brace,amplitude=3pt,mirror},black!60]
    (0,-0.62)--(2.85,-0.62);
  \node[anchor=north,font=\tiny] at (1.5,-0.71)
    {interior gaps of the marker copy};
\end{tikzpicture}
\caption{The gap map of one state window, with gap numbers measured within
the window.
The patterns $\ins_g(T)$ make the interior gaps of the marker copies
inactive.  The brace illustrates these gaps for the $S$-copy, and the same
applies to each $N$-copy.  The two gaps $P_0,P_1$ remain as sites.  Gaps $12$ and $22$ are
made inactive by $A$-based patterns.  Gaps $32$ and $42$ are made inactive
by $U$-based patterns for nonfinal windows and by $U_E$-based patterns for
the final window.}
\label{ext:fig:gapmap}
\end{figure}
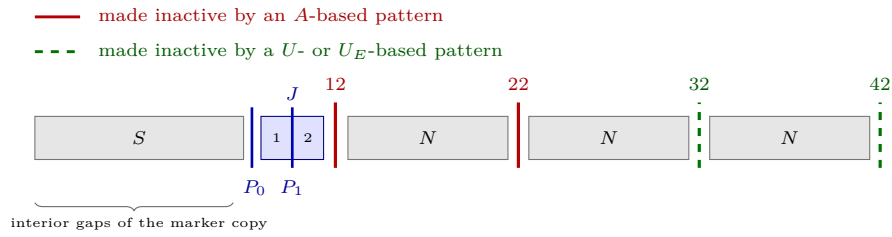

Figure~\ref{ext:fig:gapmap} summarizes how the repeated families treat the
gaps in one state window.

\begin{definition}[The state forbidden set]
\label{ext:def:bstate}
The \emph{state forbidden set} is
\[
\begin{aligned}
\Bstate={}&\mathcal B_{\mathrm{core}}
\cup\mathcal B_{\mathrm{gap}}
\cup\mathcal B_{\mathrm{cap}}
\cup\mathcal B_{\mathrm{conf}}\\
&\cup\mathcal B_{\mathrm{overlap}}
\cup\mathcal B_{\mathrm{order}}
\cup\mathcal B_{\mathrm{boundary}}.
\end{aligned}
\]
\end{definition}

\begin{proposition}[Size of $\Bstate$]
\label{ext:prop:bstate-size}
The set $\Bstate$ contains $51$ distinct permutations, consisting of $37$ permutations of length $11$, one of length
$22$, four of length $41$, six of length $43$, and three of length $44$.
\end{proposition}

\begin{proof}
The seven families in Definition~\ref{ext:def:bstate} have sizes $36$, $2$,
$2$, $1$, $2$, $1$, and $7$, giving $51$ in total and the displayed length
distribution.  A short program checks every pair of the resulting
permutations and confirms that they are distinct.
\end{proof}

\begin{table}[ht]
\centering\small
\setlength{\tabcolsep}{5pt}
\renewcommand{\arraystretch}{1.12}

\begin{tabularx}{\textwidth}{
  >{\raggedright\arraybackslash}p{2.3cm}
  >{\raggedright\arraybackslash}X
  >{\centering\arraybackslash}p{0.7cm}
  >{\raggedright\arraybackslash}p{4.5cm}
}
\toprule
Family & Generator & \# & Role \\
\midrule

$\mathcal B_{\mathrm{core}}$
  & $\ins_g(T)$, $T\in\{D,S,E,N\}$, $1\le g\le9$
  & $36$
  & Prevent insertions inside marker copies. \\

$\mathcal B_{\mathrm{gap}}$
  & $\ins_{12}(A)$, $\ins_{22}(A)$
  & $2$
  & Make gaps $12$ and $22$ of every state window inactive. \\

$\mathcal B_{\mathrm{cap}}$
  & $\ins^{\nearrow 2}_{10}(A)$, $\ins^{\nearrow 2}_{11}(A)$
  & $2$
  & Ensure that each designated site is used at most once. \\

$\mathcal B_{\mathrm{conf}}$
  & $\ins^{\nearrow}_{10,11}(A)$
  & $1$
  & Prevent both sites of a state block from being used. \\

$\mathcal B_{\mathrm{overlap}}$
  & $\ins_{10}(U)$, $\ins_{20}(U)$
  & $2$
  & Make gaps $32$ and $42$ of every nonfinal state window inactive. \\

$\mathcal B_{\mathrm{order}}$
  & $4312[1,1,N,E]$
  & $1$
  & Force inserted maxima to increase from left to right. \\

$\mathcal B_{\mathrm{boundary}}$
  & $\ins_0(D)$ and the six insertion patterns based on $U_D,U_E$
  & $7$
  & Make the gaps at the two ends of the sequence of state windows inactive. \\

\midrule
\multicolumn{2}{r}{Total}
  & $51$
  & \\
\bottomrule
\end{tabularx}
\caption{The seven families of patterns in $\Bstate$ and what each family
enforces.}
\label{ext:tab:categories}
\end{table}

\subsection{Routing the variable occurrences}
The base permutation supplies the state blocks that will be assigned bits.  We now
determine how many are needed and how they must be connected to encode the
formula.  We arrange them into levels.

Treat each appearance of a variable in $\Phi$ as a distinct variable
occurrence and initially order the $h$ variable occurrences clause by clause.
Relabel the variables
$x_1,\dots,x_n$, choosing $x_n$ to be the variable of the last
variable occurrence in this order, and put
$
  \lambda=\lceil\log_2 n\rceil.
$

Levels $0,\dots,\lambda+1$ each contain one state block for every variable
occurrence.  The blocks representing the same variable occurrence in
successive levels are required
to have the same bit, so that bit is preserved while its position may
change.  Levels $0$ and $1$ both use clause order, where each consecutive triple represents one clause. The
next $\lambda$ levels gradually change clause order into an order in which all
variable occurrences belonging to each variable are consecutive.  The final
level $\lambda+2$ is
the merge level.  It contains one state block for each variable, whose bit is
required to agree with the bits of all its variable occurrences.  We write
$b_c$ for the bit assigned to state block $c$.

The intermediate levels are needed because the proof of correctness
requires that, when a new level is read from left to right, the positions of
its variable occurrences in the preceding level have longest decreasing
subsequence at most $2$.  A direct change from clause order to variable order need not
satisfy this requirement.  We instead pass through a sequence of stable
binary partitions.  Each pass preserves the order within the two groups.  The
preceding-level positions therefore form two increasing subsequences and
satisfy the requirement.

Give $x_j$, for $1\le j<n$, the $\lambda$-bit code representing $j-1$.
Give $x_n$ the all-ones code $2^\lambda-1$.
Starting from clause order, for $0\le a<\lambda$ stably partition the
variable occurrences according to bit position $a$ of their codes,
from least significant to most significant. All variable occurrences whose
code bit is $0$ come first, followed by those whose code bit is $1$,
while the relative order within each group is preserved. After pass $a$ the
occurrences are ordered by the lowest $a+1$ code bits. After
$\lambda$ passes they are grouped in variable order
$x_1,x_2,\dots,x_n$ 
with all variable occurrences belonging to each variable consecutive. The
last variable occurrence has code bit $1$ in every pass and therefore
remains last throughout the reordering.

\subsubsection{Levels, wires, and clause triples}
The following construction turns these orders into levels of state
blocks. It specifies which pairs of blocks must have the same bit and which
triples must satisfy the not-all-equal condition.

Set $r=(\lambda+2)h+n$ 
and, for $0\le\ell\le\lambda+1$ and $0\le i<h$, index the state blocks of
$\rho^{\star}_r$ by $c(\ell,i)=\ell h+i$. Thus,  $c(\ell,i)$ is the index of the state block
$F(\ell,i+1)$ from the main text. 
For each $\ell$, the blocks
$c(\ell,0),\dots,c(\ell,h-1)$ form level $\ell$.
The final $n$ state blocks, with indices
\[
  (\lambda+2)h+j-1,
  \qquad 1\le j\le n,
\]
form level $\lambda+2$, with block
$(\lambda+2)h+j-1$ corresponding to $x_j$.

A \emph{wire} is a directed pair $s\to t$ connecting a state block $s$ in
one level to a state block $t$ in the following level.
We call $s$ its \emph{source} and $t$ its \emph{target}.
A wire requires $b_s=b_t$.  A \emph{transition} is the set of wires between
two successive levels.  Let $\mathcal W_\Phi$ be the following set of wires:
\begin{enumerate}[label=(\roman*)]
\item In the transition from level $0$ to level $1$, add
      $c(0,i)\to c(1,i)$ for every $i$.  The triples
      \[
        \bigl(c(1,3k),c(1,3k+1),c(1,3k+2)\bigr),
        \qquad 0\le k<q,
      \]
      form the set $\mathcal C_\Phi$ of \emph{clause triples}.
\item For $0\le a<\lambda$, the transition from level $a+1$ to level $a+2$
      performs the $a$th stable binary partition described above.  Add a wire
      from each variable occurrence in level $a+1$ to the block carrying the
      same variable occurrence in level $a+2$.
\item In the transition from level $\lambda+1$ to level $\lambda+2$, if a
      variable occurrence of $x_j$ occupies position $i$ in level
      $\lambda+1$, add
      \[
        c(\lambda+1,i)\to (\lambda+2)h+j-1.
      \]
\end{enumerate}
We call the transition from level $0$ to level $1$ the \emph{clause
transition}, and level $1$ the \emph{clause level}.  We call level
$\lambda+2$ the \emph{merge level}, and its state blocks the \emph{merge
blocks}.  The final transition is the \emph{merge transition}.

\begin{lemma}[Wire separation]
\label{ext:lem:wire-separation}
Every wire $s\to t$ satisfies $t-s\ge2$.
\end{lemma}

\begin{proof}
First consider a wire between two levels that each contain $h$ blocks.  Suppose
its source is at position $i$ in level $\ell$ and its target is at position
$j$ in level $\ell+1$.  The definitions of the block indices give
\[
  s=\ell h+i,
  \qquad
  t=(\ell+1)h+j,
\]
and hence
\[
  t-s=h+j-i.
\]
In the clause transition, the source and target have the same position, so
$j=i$ and $t-s=h\ge3$.  In a reordering transition, if $i\le h-2$, then
$j\ge0$ gives
\[
  t-s=h+j-i\ge h-i\ge2.
\]
If $i=h-1$, the source is the last variable occurrence.  This variable
occurrence remains last through every stable binary partition, so $j=h-1$
and $t-s=h$.

It remains to consider the merge transition.  If the source is at position
$i$ and its variable is $x_j$, then
\[
  s=(\lambda+1)h+i,
  \qquad
  t=(\lambda+2)h+j-1,
\]
so $t-s=h+j-1-i$.  If $i\le h-2$, then $t-s\ge h-i\ge2$.  If $i=h-1$,
the source is the final variable occurrence, which belongs to $x_n$.
Therefore $j=n$ and $t-s=n\ge2$.
\end{proof}

Figure~\ref{fig:routing} summarizes the levels and the transitions.

\begin{lemma}[Correctness of the routing]
\label{ext:lem:equivalence}
The formula $\Phi$ is NAE-satisfiable if and only if bits can be assigned to
the $r$ state blocks so that every wire has equal endpoint bits and every
clause triple is nonconstant.
\end{lemma}

\begin{proof}
Suppose $\Phi$ has an NAE-satisfying assignment $\beta$.  Give each level-$0$
block the truth value of the variable it represents, and propagate these bits
through the clause and reordering levels.  Give the merge block for $x_j$ the
bit $\beta(x_j)$.  Every wire has equal endpoint bits, and each clause triple
contains the truth values of the three variable occurrences in one clause, so
it is nonconstant.

Conversely, suppose the block bits satisfy every wire and clause triple.  The
clause and reordering transitions propagate the bit associated with each
variable occurrence while merely changing its position.  The merge wires
therefore force all variable occurrences belonging to $x_j$ to have the bit
of its merge block.  Define $\beta(x_j)$ to be this bit.
Each clause triple then has the three truth values assigned to its clause and is
nonconstant.  Thus $\beta$ NAE-satisfies $\Phi$.
\end{proof}

\subsection{Encoding the wires and clauses}
\label{ext:sec:decoration}

We now augment $\rho^{\star}_r$ to encode the wires and clause triples.  Each
wire $s\to t$ is encoded by two patterns whose values identify the source $s$
and whose positions bracket the target block $F(t)$.  Each clause triple is
encoded by two further patterns placed on either side of its three state
blocks.

\begin{definition}[Constraint permutations]
\label{ext:def:constraint-permutations}
Let
\[
  Q_L,Q_R,C_L,C_R\in\mathfrak S_{10}
\]
be the four permutations listed and plotted in
Appendix~\ref{ext:app:fixed-permutation-data}.  Collectively they are the \emph{constraint
permutations}. These four permutations identify the wires and clause triples.
\end{definition}

\begin{proposition}[Properties of the constraint permutations]
\label{ext:prop:constraint-permutation-facts}
The four constraint permutations are pairwise distinct and simple, none is
equal to a marker permutation, and every constraint permutation $T$ satisfies
$\LDS(T)=6$.
\end{proposition}

\begin{proof}
A short program exhaustively checks the four displayed permutations of length
$10$, together with the four marker permutations, and confirms the stated
properties.
\end{proof}

The designated copies of $Q_L$, $Q_R$, $C_L$, and $C_R$ are
collectively called \emph{constraint copies}.  These copies encode the wires and clause
triples.  The four violation
patterns introduced in the next section exclude bit assignments with unequal
wire endpoints or constant clause triples.
We first specify the available relative value positions for these
copies.

\begin{definition}[Value band]
\label{ext:def:band}
Let $s$ be the source block of a wire.  In the value interval between $S(s)$
and $J(s)$, reserve four possible \emph{slots}, ordered from lowest to highest as the
$C_L$-, $Q_L$-, $Q_R$-, and $C_R$-slots of $s$.
\[
  S(s)
  \;<\;\text{$C_L$-slot}
  \;<\;\text{$Q_L$-slot}
  \;<\;\text{$Q_R$-slot}
  \;<\;\text{$C_R$-slot}
  \;<\;J(s),
\]
where $U<V$ means that every value in $U$ is smaller than every value in $V$.
An occupied slot contains one copy of the corresponding constraint
permutation and consists of ten consecutive values.  An unused slot contains
no entries.  The union of the occupied slots is the \emph{value band of $s$}.
\end{definition}

For each wire $s\to t$, copies of $Q_L$ and $Q_R$ occupy the
value band of $s$ and lie on opposite sides of $F(t)$ in position.
Thus their values identify the source, while their positions identify the
target. Similarly, copies of $C_L$ and $C_R$ bracket the three state
blocks of a clause triple. We next give the precise placement rules.
Every block that is a source of a wire has exactly one outgoing wire,
so the corresponding copies can be indexed by that source.
The permutation $\rho_\Phi$ is obtained from
$\rho^{\star}_r$ by the following placement rules:

\begin{enumerate}[label=(D\arabic*), ref=D\arabic*]
\item\label{ext:D1}
For each state block $t$ with incoming wires
$s_1\to t,\dots,s_m\to t$, where 

$s_1<\cdots<s_m$, add two constraint copies
$Q_L(s_i)$ and $Q_R(s_i)$ for each $i$.  Place
\[
  Q_L(s_1),\dots,Q_L(s_m)
\]
consecutively immediately before $F(t)$, and place
\[
  Q_R(s_1),\dots,Q_R(s_m)
\]
consecutively immediately after $F(t)$.  For each $i$, the values of
$Q_L(s_i)$ and $Q_R(s_i)$ occupy the $Q_L$- and $Q_R$-slots, respectively,
in the value band of $s_i$.

\item\label{ext:D2}
For each clause triple $(t_0,t_1,t_2)\in\mathcal C_\Phi$, let
$s_0\to t_0$ be the incoming wire to $t_0$. Add two constraint copies 
$C_L(s_0)$ and $C_R(s_0)$.
Place $C_L(s_0)$ immediately before the $Q_L$-copy at $t_0$ and
$C_R(s_0)$ immediately after the $Q_R$-copy at $t_2$.  The values of
$C_L(s_0)$ and $C_R(s_0)$ occupy the $C_L$- and $C_R$-slots, respectively,
in the value band of $s_0$.

\item\label{ext:D3}
No other constraint copies are added.
\end{enumerate}

Rules~\ref{ext:D1}--\ref{ext:D3}, together with the value order in
Definition~\ref{ext:def:band}, determine $\rho_\Phi$ uniquely.

\begin{lemma}[Separation of nonadjacent windows]
\label{ext:lem:envelope}
Let $c$ and $c'$ index state blocks of $\rho_\Phi$, and suppose that
$c'\ge c+2$.  Then the window of $c$ lies below the window of $c'$.  If $c$
is a source block, its value band also lies below the window of $c'$.

In particular, for every wire $s\to t$, the window and value band of $s$ lie
below the window of $t$.
\end{lemma}

\begin{proof}
By Definition~\ref{ext:def:oscillation}, the defining-sequence values
underlying the window of $c$ are
\[
  (a_{4c+3},\dots,a_{4c+6})
  =(4c+6,\;4c+3,\;4c+8,\;4c+5),
\]
whose maximum is $4c+8$.  The minimum of the corresponding four values for
$c'$ is $4c'+3$.  Since $c'\ge c+2$,
\[
  4c'+3\ge4c+11>4c+8.
\]
Standardization and inflation preserve this order, so the window of $c$ lies
below the window of $c'$.  If $c$ is a source block, its value band lies below
$J(c)$, which belongs to the window of $c$, and hence also lies below the
window of $c'$.

For a wire $s\to t$, Lemma~\ref{ext:lem:wire-separation} gives $t\ge s+2$,
so the preceding result applies with $c=s$ and $c'=t$.
\end{proof}

The placement rules also determine which state blocks lie between the
constraint copies.

\begin{lemma}[Positional brackets]
\label{ext:lem:bracket}
For every wire $s\to t$, the only state block between $Q_L(s)$ and $Q_R(s)$
in position is $F(t)$, and no other marker copy lies between them.

For every clause triple $(t_0,t_1,t_2)\in\mathcal C_\Phi$, let
$s_0\to t_0$ be the incoming wire to $t_0$.  The state blocks between
$C_L(s_0)$ and $C_R(s_0)$ in position are exactly
$F(t_0),F(t_1),F(t_2)$, in that order.
\end{lemma}

\begin{proof}
Immediate from Rules~\ref{ext:D1}--\ref{ext:D3} and the definition of
$\mathcal C_\Phi$.
\end{proof}

The construction uses $(\lambda+2)h$ wires and $q$ clause triples.  Consequently
\begin{equation}
\label{ext:eq:size}
  |\rho_\Phi|=(42r+40)+20(\lambda+2)h+20q.
\end{equation}
The reduction maps $\Phi$ to $(\rho_\Phi,1^r)$.  Since $n\le h$, we have
$\lambda=O(\log h)$ and $r=O(h\log h)$.  Moreover, $q=h/3$, so
Equation~\eqref{ext:eq:size} gives $|\rho_\Phi|=O(h\log h)$.  Thus the output
has polynomial size and can be constructed in polynomial time.

\subsection{The four violation patterns}
\label{ext:sec:violation-patterns}

The constraint copies identify the wires and clause triples but do not yet
restrict the bits assigned to their state blocks.  We now define four forbidden
patterns that complete the construction.  Two forbid unequal endpoints of a
wire, and two forbid the constant assignments on a clause triple.

\subsubsection{Equality along wires}
\label{ext:sec:equality}

\begin{definition}[$\Aedge$]
\label{ext:def:Aeq}
Let $s\to t\in\mathcal W_\Phi$.  Define $\Aedge$ to be the
permutation induced by the following $104$ entries, given in positional order:
\begin{enumerate}[label=(\roman*)]
\item the four blocks of the window of $s$ \hfill($42$ entries).
\item the copy $Q_L(s)$ \hfill($10$).
\item the state block $F(t)$ \hfill($12$).
\item the copy $Q_R(s)$ \hfill($10$).
\item the three $N$-copies in the window of $t$ \hfill($30$).
\end{enumerate}
\end{definition}

\begin{lemma}[$\Aedge$ is well defined]
\label{ext:lem:Aeq-welldef}
$\Aedge$ is a permutation of length $104$, independent of $\Phi$ and of the
chosen wire.
\end{lemma}

\begin{proof}
Definition~\ref{ext:def:Aeq} fixes the positional order of the selected
entries.  The windows of $s$ and $t$ both induce $A$.  By
Lemma~\ref{ext:lem:envelope}, the window and value band of $s$ lie below the
window of $t$.  By Definition~\ref{ext:def:band} and Rule~\ref{ext:D1},
$Q_L(s)$ and $Q_R(s)$ lie between $S(s)$ and $J(s)$ in value, with
$Q_L(s)<Q_R(s)$.  Thus the induced permutation is independent of $\Phi$ and
of the chosen wire.  Its length is $42+10+12+10+30=104$.
\end{proof}

In $\Aedge$ the source state block occupies positions $1,\dots,12$, so its
sites are gaps $10$ and $11$.  The target state block occupies positions
$53,\dots,64$, so its sites are gaps $62$ and $63$.  We therefore define
\[
  \Pneq{01}=\ins^{\nearrow}_{10,63}(\Aedge),
  \qquad
  \Pneq{10}=\ins^{\nearrow}_{11,62}(\Aedge).
\]
Both patterns have length $106$.  The first represents state $0$ at the source
and state $1$ at the target.  The second represents state $1$ at the source
and state $0$ at the target.

\subsubsection{Not-all-equal clauses}
\label{ext:sec:nae}

\begin{definition}[$\Anae$]
\label{ext:def:Anae}
Let $(t_0,t_1,t_2)\in\mathcal C_\Phi$, and let $s_0\to t_0$ be the incoming
wire to $t_0$.  Define $\Anae$ to be the pattern induced by the following
$188$ entries, given in positional order:
\begin{enumerate}[label=(\roman*)]
\item the four blocks of the window of $s_0$ \hfill\textnormal{(42)}.
\item the copy $C_L(s_0)$ \hfill\textnormal{(10)}.
\item the four blocks of the window of $t_0$ \hfill\textnormal{(42)}.
\item the four blocks of the window of $t_1$ \hfill\textnormal{(42)}.
\item the state block $F(t_2)$ \hfill\textnormal{(12)}.
\item the copy $C_R(s_0)$ \hfill\textnormal{(10)}.
\item the three $N$-copies in the window of $t_2$ \hfill\textnormal{(30)}.
\end{enumerate}
\end{definition}

\begin{lemma}[$\Anae$ is well defined]
\label{ext:lem:Anae-welldef}
$\Anae$ is a permutation of length $188$, independent of $\Phi$ and of the
chosen clause triple.
\end{lemma}

\begin{proof}
Definition~\ref{ext:def:Anae} fixes the positional order of the selected
entries.  By Definition~\ref{ext:def:band} and Rule~\ref{ext:D2}, the copies of $C_L$ and $C_R$
have fixed positions in the value band of $s_0$.  By
Lemma~\ref{ext:lem:envelope}, the source window and its value band lie below
the three target windows.  It remains only to determine the relative values
of those target windows.

Write $t_0=u$, $t_1=u+1$, and $t_2=u+2$.  The defining-sequence values
underlying the window of a state block $v$ are
\[
  4v+6,\;4v+3,\;4v+8,\;4v+5.
\]
Hence the values underlying the three target windows are
$4u+\{3,5,6,8\}$, $4u+\{7,9,10,12\}$, and
$4u+\{11,13,14,16\}$.  Their increasing order is
\[
  \underbrace{3,5,6}_{t_0},\underbrace{7}_{t_1},
  \underbrace{8}_{t_0},\underbrace{9,10}_{t_1},
  \underbrace{11}_{t_2},\underbrace{12}_{t_1},
  \underbrace{13,14,16}_{t_2},
\]
up to the common offset $4u$.  This order is independent of $u$, and every
inflated block induces a fixed permutation.  Thus $\Anae$ is independent of
$\Phi$ and of the chosen clause triple.  Its length is
$42+10+42+42+12+10+30=188$.
\end{proof}

In $\Anae$ the state blocks of $t_0,t_1,t_2$ occupy positions
$53$--$64$, $95$--$106$ and $137$--$148$, so their sites are the gaps
\[
  P_0: 62,\,104,\,146, \qquad P_1: 63,\,105,\,147.
\]
We therefore define
\[
  \Pmono{000}=\ins^{\nearrow}_{62,104,146}(\Anae),
  \qquad
  \Pmono{111}=\ins^{\nearrow}_{63,105,147}(\Anae).
\]
Both patterns have length $191$.  In the two patterns, the inserted maxima
occupy, respectively, the three $P_0$ sites and the three $P_1$ sites of the
state blocks in the clause triple.

Collectively, these four permutations form the set of \emph{violation patterns}
\[
  \mathcal B_{\mathrm{viol}}
  =\{\Pneq{01},\Pneq{10},\Pmono{000},\Pmono{111}\}.
\]

Table~\ref{ext:tab:constraint-effects} shows the assignments excluded by the
violation patterns.

\begin{table}[ht]
\centering
\small
\renewcommand{\arraystretch}{1.15}
\begin{tabularx}{\textwidth}{@{}l X X@{}}
\toprule
constraint & satisfying condition & forbidden patterns \\
\midrule
wire $s\to t$
  & $b_s=b_t$
  & $\Pneq{01}$ forbids $(b_s,b_t)=(0,1)$, while
    $\Pneq{10}$ forbids $(b_s,b_t)=(1,0)$ \\
\addlinespace
clause triple $(t_0,t_1,t_2)$
  & $(b_{t_0},b_{t_1},b_{t_2})$ is not constant
  & $\Pmono{000}$ forbids $(0,0,0)$, while
    $\Pmono{111}$ forbids $(1,1,1)$ \\
\bottomrule
\end{tabularx}
\caption{The equality and NAE conditions enforced by the violation patterns.}
\label{ext:tab:constraint-effects}
\end{table}

\subsection{The forbidden set \texorpdfstring{$\Bfinal$}{Bfinal}}
\label{ext:sec:final-set}

The patterns in $\Bstate$ control where maxima may be inserted.  The four
violation patterns impose the Boolean conditions on the wires and clause
triples.  Their union is the
fixed forbidden set used in the reduction.

\begin{definition}[$\Bfinal$]
\label{ext:def:bfinal}
$\Bfinal=\Bstate\cup\mathcal B_{\mathrm{viol}}$.
\end{definition}



The reduction and the forbidden set $\Bfinal$ are now fully defined.  The rest of the
appendix proves that the base permutation has extension depth $r$ and that every
maximum-depth extension assigns a unique bit to each
state block.  It then proves that adding constraint copies preserves both
conclusions and that every occurrence of a violation pattern corresponds to an
unequal wire or a constant clause triple.

\subsection{Verification of the base permutation and $\rho_\Phi$}
\label{ext:sec:encoding-proof}

We first prove that every occurrence of a marker or constraint permutation is
one of the copies placed above.  We then
show that $\Bstate$ leaves exactly two sites in every state block and gives
depth $r$, both before and after the constraint copies are added.  Finally,
we prove that the violation patterns occur exactly on wires with unequal
endpoint bits and clause triples with equal bits.

\subsubsection{Occurrences of \texorpdfstring{$\vartheta$}{theta} in increasing oscillations}
\label{ext:sec:oscillation-occurrences}

This subsection proves the properties of increasing oscillations used later.

\begin{lemma}[Occurrences of $\vartheta$ in $\omega_m$]
\label{ext:lem:osc-window}
Let $m\ge4$.  The occurrences of $\vartheta$ in $\omega_m$ are exactly the
sets of four consecutive positions
\[
  p,p+1,p+2,p+3,
\]
where $p$ is odd and $p+3\le m$.
\end{lemma}

\begin{proof}
Write $a_1,a_2,\dots$ for the defining sequence of
Definition~\ref{ext:def:oscillation}, so $a_{2k-1}=2k+2$ and $a_{2k}=2k-1$.
Let $G_m$ be the inversion graph on positions $1,\dots,m$, with an edge
$\{i,j\}$ for $i<j$ when $a_i>a_j$.  Directly from the formulas for $a_i$,
\[
  E(G_m)
  =\bigl\{\{2k-1,2k\}:2k\le m\bigr\}
   \cup
   \bigl\{\{2k-1,2k+2\}:2k+2\le m\bigr\}.
\]
Here $E(G_m)$ denotes the edge set of $G_m$.
If $m$ is even, $G_m$ is the path
\[
  2-1-4-3-6-5-\cdots-m-(m-1).
\]
If $m$ is odd, the same path uses positions $1,\dots,m-1$, and position $m$
is isolated.

Write the vertices in their path order as
\[
  v_{2i-1}=2i,\qquad v_{2i}=2i-1
  \quad\text{for }1\le i\le\lfloor m/2\rfloor.
\]
The inversion graph of $\vartheta=3142$ has edges
\[
  \{1,2\},\ \{1,4\},\ \{3,4\},
\]
and is the path $2-1-4-3$.  Hence the positions of any occurrence of
$\vartheta$ induce a four-vertex path in $G_m$.  Since $G_m$ is itself a path
apart from a possible isolated vertex, these positions must be
$v_s,v_{s+1},v_{s+2},v_{s+3}$ for some $s$.

If $s$ is odd, write $s=2i-1$.  The four vertices are
\[
  v_{2i-1},\dots,v_{2i+2}
  =2i,\,2i-1,\,2i+2,\,2i+1,
\]
which, in increasing position order, are $2i-1,\dots,2i+2$.  Moreover,
\[
  \st(a_{2i-1},\dots,a_{2i+2})=3142=\vartheta.
\]
If $s$ is even, write $s=2i$.  The selected positions in increasing order are
\[
  2i-1,\ 2i+1,\ 2i+2,\ 2i+4,
\]
and
\[
  \st(a_{2i-1},a_{2i+1},a_{2i+2},a_{2i+4})
  =2413\ne\vartheta.
\]
Thus every occurrence of $\vartheta$ occupies four consecutive positions
beginning at an odd position.  Conversely, the displayed standardization
shows that every such set of positions induces $\vartheta$.
\end{proof}

We use the following bound for decreasing subsequences.

\begin{lemma}[Subadditivity of $\LDS$]
\label{ext:lem:lds-union}
If the entries  of $\pi$ are partitioned into parts $P_1,\dots,P_t$ with
$\LDS(P_i)\le a_i$, then $\LDS(\pi)\le a_1+\cdots+a_t$.
\end{lemma}

\begin{proof}
A decreasing subsequence meets $P_i$ in a decreasing subsequence, hence in at
most $a_i$ entries.  Summing over the parts gives the stated bound.
\end{proof}

\begin{lemma}[Longest decreasing subsequences of $\omega_m$]
\label{ext:lem:eta-lds}
$\LDS(\omega_m)\le2$ for all $m\ge2$.
\end{lemma}

\begin{proof}
In the defining sequence, the values in odd positions form the increasing
sequence $4,6,8,\dots$, while the values in even positions form the increasing
sequence $1,3,5,\dots$.  Standardization preserves these orders, so
$\omega_m$ is the union of two increasing subsequences and
$\LDS(\omega_m)\le2$ by Lemma~\ref{ext:lem:lds-union}.
\end{proof}

We next prove that an occurrence of a marker or constraint permutation cannot
use entries  from several copies placed by the construction.

\subsubsection{Occurrences of marker and constraint permutations}
\label{ext:sec:isolation}

The following lemma shows how a simple occurrence can meet a block.

\begin{lemma}[A simple occurrence meeting a block]
\label{ext:lem:simple-interval}
Let $\tau$ be a simple permutation, and let $I$ be a block of a permutation
$\sigma$. If an occurrence of $\tau$ in $\sigma$ meets $I$ in at least two
entries, then the whole occurrence lies inside $I$.
\end{lemma}

\begin{proof}
Let $O$ be an occurrence of $\tau$ in $\sigma$.  We show that $O\cap I$ is a block of the permutation induced by $O$.  Since $I$ occupies consecutive
positions in $\sigma$, the entries  of $O\cap I$ occupy consecutive positions
within $O$.  Similarly, any entry of $O$ whose value lies between the values of
two entries  in $O\cap I$ must also lie in $I$, because $I$ contains a consecutive
set of values.  Hence $O\cap I$ occupies consecutive value ranks within $O$.
Therefore $O\cap I$ is a block of $O$.  Since $\tau$ is simple and
$|O\cap I|\ge2$, this block must be all of $O$, and hence $O\subseteq I$.
\end{proof}

Throughout this subsection, a \emph{configuration} means either the base
permutation $\rho^{\star}_r$ or $\rho_\Phi$, together with any
inserted maxima placed at sites.

Here \emph{marker copy} and \emph{constraint copy} refer to copies placed by
the construction, whereas \emph{occurrence} has its pattern-containment
meaning.  We prove below that every occurrence of one of the eight marker and
constraint permutations is a copy placed by the construction.

\paragraph{Blocks occupied by construction copies}

We first show that every placed marker or constraint copy remains a block
after constraint copies and inserted maxima are added.  This lets us use
simplicity whenever an occurrence meets one copy more than once.

\begin{lemma}[Marker and constraint copies are blocks]
\label{ext:lem:marker-intervals}
In every configuration, each marker copy and each constraint copy
is a block.
\end{lemma}

\begin{proof}
\emph{Positions.}  Each copy was placed as a run of consecutive positions.
By~\ref{ext:D1}--\ref{ext:D3}, constraint copies are placed immediately
before or after state blocks or adjacent runs of constraint copies, never
inside a marker copy.  Inserted maxima lie only at sites, which occur after
the marker copy inside a state block.  Hence no later placement separates the
positions of a marker or constraint copy, so every such copy still occupies
consecutive positions.

\emph{Values.}  A marker copy outside a state block occupies the entire value
interval of its corresponding entry of $\omega_{4r+4}$, which is consecutive.
Inside a state block, the marker copy $S(c)$ occupies the bottom $10$ values
of the block's value interval.  If $c$ is a source block,
Definition~\ref{ext:def:band} places its value band strictly between $S(c)$
and $J(c)$.  Thus every marker copy occupies consecutive values.  Each
constraint copy fills the corresponding consecutive $10$-value slot from
Definition~\ref{ext:def:band}.  Every inserted maximum is greater than every
root entry, so it cannot separate the consecutive values of a marker or
constraint copy.  Hence every marker and constraint copy still occupies
consecutive values.
\end{proof}

State blocks themselves need not be blocks in the formal sense.  A state block $F(s)$ of
$\rho_\Phi$ is a run of consecutive positions,
but as soon as the value band of $s$ is nonempty its values are \emph{not}
consecutive: the band separates $S(s)$ from $J(s)$ in value while the band's
entries  occur outside $F(s)$ in position.  Lemma~\ref{ext:lem:simple-interval} is therefore
applied below to marker and constraint copies, never to state blocks.

\begin{lemma}[Inserted maxima increase from left to right]
\label{ext:lem:ordering}
Let $\pi$ be a configuration avoiding $\Bstate$.  Then its inserted maxima
increase from left to right in value.
\end{lemma}

\begin{proof}
In $\rho^{\star}_r$, the third $N$-copy following the final state block and
the final copy of $E$ lie to the right of every site, and every value in that
$N$-copy is smaller than every value in $E$.  Constructing $\rho_\Phi$ and
inserting maxima neither deletes nor reorders root entries.  Suppose inserted
maxima $x$ and $y$ satisfy $x<y$ in position but $x>y$ in value.  Then $x,y$
together with these copies of $N$ and $E$ form an occurrence of
$4312[1,1,N,E]$, contradicting the assumed
$\Bstate$-avoidance.
\end{proof}

\paragraph{Decreasing subsequences among constraint copies}

We next bound how many distinct constraint copies can contribute to a
decreasing subsequence.  We will combine this bound with the common
decreasing-subsequence length of the fixed marker permutations.

\begin{lemma}[Constraint copies have $\LDS$ at most $2$]
\label{ext:lem:aux-lds}
Let $\beta$ be the pattern induced by choosing one arbitrary entry from each
constraint copy of $\rho_\Phi$.  Then $\LDS(\beta)\le2$.
\end{lemma}

\begin{proof}
By Lemma~\ref{ext:lem:marker-intervals}, each constraint copy is a block,
so the relative order of the chosen entries  depends only on their copies.  Let
$\delta$ be a decreasing subsequence of $\beta$.  Associate the two copies of
each wire with its transition, and associate every copy of $C_L$ or $C_R$ with the
clause transition.

If $0\le a<b\le\lambda+1$, every copy associated with the transition out of
level $a$ lies before and below every copy associated with the transition
out of level $b$.  Indeed, their positions surround blocks in the
corresponding target levels, and their values lie in bands of the
corresponding source levels.  Hence $\delta$ uses copies from only one
transition.

\emph{The clause transition.}  For a clause triple with consecutive sources
$s_0,s_1,s_2$, its eight constraint copies appear in positional order
\[
  C_L(s_0),Q_L(s_0),Q_R(s_0),Q_L(s_1),Q_R(s_1),
  Q_L(s_2),Q_R(s_2),C_R(s_0),
\]
and their value ranks are
\[
  1,2,3,5,6,7,8,4.
\]
Distinct triples increase in both position and value, so $\delta$ uses copies
from only one triple.  The first seven ranks are increasing, so $\delta$
contains at most one of them and possibly the final rank.  Thus
$|\delta|\le2$.

\emph{A reordering transition.}  Reading the target level from left to right,
the source indices appear first for the variable occurrences whose relevant
binary bit is $0$, in their original order, and then for those whose bit is
$1$, also in their original order.  Thus the source indices are the union of two increasing
subsequences.  For each source, the selected entries  from its $Q_L$- and
$Q_R$-copies form an increasing pair because the $Q_L$-slot precedes the
$Q_R$-slot in value.  Hence $\delta$ contains at most one of them.  Since each
value band lies between $S(s)$ and $J(s)$, the value bands increase with the
source index.  The source indices used by $\delta$ therefore form a decreasing
subsequence of the displayed target order, so $|\delta|\le2$.

\emph{The merge transition.}  At each merge block, the incoming $Q_L$-copies
form one run in increasing source order before the block, and the corresponding
$Q_R$-copies form one run in increasing source order after it.  Each run is
increasing in both position and value.  Different merge blocks increase in
both position and source value, so $\delta$ uses copies from only one merge
block and at most one copy from each of its two runs.  Thus $|\delta|\le2$.

In every case $|\delta|\le2$, and therefore $\LDS(\beta)\le2$.
\end{proof}

\paragraph{Occurrences are construction copies}

We now combine the block and decreasing-subsequence bounds.  An occurrence
meeting several marker or constraint copies would have a decreasing
subsequence that is too short.  Thus every occurrence must lie in a single
such copy.

\begin{lemma}[Decreasing subsequences meeting each copy once]
\label{ext:lem:lds-budget}
Let $\beta$ be the pattern induced by choosing entries  from a configuration
$\pi$ whose inserted maxima increase from left to right, with at most one
entry chosen from each marker copy and each constraint copy.  Then
$\LDS(\beta)\le5$.
\end{lemma}

\begin{proof}
Let $\delta$ be a decreasing subsequence of $\beta$.

First consider the entries  of $\delta$ belonging to the base permutation
$\rho^{\star}_r$.  Each such entry lies in a block corresponding to one entry
of $\omega_{4r+4}$.  If that block is a marker copy, at most one entry was
chosen from it.  If it is a state block, its selected entry from
$S$, if any, and all its selected entries  from $J$ form an increasing
sequence because $S$ lies below and to the left of $J$.  Since $\delta$ is
decreasing, it contains at most one entry from the whole state block.  Thus
each block corresponding to an entry of $\omega_{4r+4}$ contributes at
most one entry to $\delta$.  Replacing each such entry by the corresponding
entry of $\omega_{4r+4}$ preserves relative order, so the resulting entries  form a
decreasing subsequence of $\omega_{4r+4}$.  Lemma~\ref{ext:lem:eta-lds}
therefore bounds their number by $2$.

Likewise, at most one entry was chosen from each constraint copy.
Lemma~\ref{ext:lem:aux-lds} therefore bounds the number of constraint-copy
entries  in $\delta$ by $2$.  Finally, the inserted maxima increase from left to
right, so $\delta$ contains at most one inserted maximum.  Hence
\[
  |\delta|\le2+2+1=5. \qedhere 
\]
\end{proof}

\begin{lemma}[Occurrences of marker and constraint permutations]
\label{ext:lem:isolation}
Let $\pi$ be a configuration whose inserted maxima increase from left to right.
For every marker or constraint permutation $T$, every occurrence of $T$ in
$\pi$ is one of the copies of $T$ placed by the construction.
\end{lemma}

\begin{proof}
Let $O$ be an occurrence of $T$.  Suppose $O$ is not contained in a single
marker or constraint copy.  All such copies are blocks
by Lemma~\ref{ext:lem:marker-intervals}, and
$T$ is simple of length $10\ge4$ by
Propositions~\ref{ext:prop:marker-facts}
and~\ref{ext:prop:constraint-permutation-facts}.  Thus
Lemma~\ref{ext:lem:simple-interval} shows that $O$ meets each marker or constraint
copy in at most one entry.  Lemma~\ref{ext:lem:lds-budget} then
gives $\LDS(O)\le5$.  But $O$ induces $T$, and $\LDS(T)=6$ by
the same two propositions.  This contradicts $\LDS(O)=\LDS(T)$.

Hence $O$ lies inside a single marker or constraint copy $M$.  Every such copy
has $10$ entries, so $O=M$.  The eight permutations are pairwise distinct by
Propositions~\ref{ext:prop:marker-facts}
and~\ref{ext:prop:constraint-permutation-facts}, so $M$ is a copy of $T$ placed
by the construction.
\end{proof}

\paragraph{Occurrences of the window patterns}
We next use the marker copies to locate the windows.

\begin{lemma}[Occurrences of the state window]
\label{ext:lem:window-occurrences}
Let $\pi$ be a configuration whose inserted maxima increase from left to right.
Every occurrence of $A$ in $\pi$ is exactly the window of a state block.
\end{lemma}

\begin{proof}
Consider an occurrence of $A=\vartheta[F,N,N,N]$.  It contains one occurrence
of $S$ inside its $F$-block and three occurrences of $N$, so by
Lemma~\ref{ext:lem:isolation} these are four marker copies.  Associate each
marker copy with the entry of $\omega_{4r+4}$ whose inflated block contains it.
For the copy of $S$, this is the entry inflated by its state block.  The four
associated entries  induce $\vartheta$, so Lemma~\ref{ext:lem:osc-window} makes
them four consecutive positions beginning at an odd position.  The selected
$S$-copy lies in the first inflated block and belongs to a state block.
Hence the four consecutive positions begin with that state block and form
its window.  Denote this state block by $t$.

It remains to identify the two entries  playing $J$.  They lie positionally
between the selected copy of $S$ and the first selected copy of $N$, with
values above $S$ and below the second following $N$-copy.  The candidates in
that positional range are the copy of $J$ in the state block, maxima inserted
at its sites, the copies $Q_R(s)$ for wires $s\to t$, and possibly a copy
$C_R(s)$ when $t$ is the last block of a clause triple.  Inserted maxima lie
above every root entry.  In the first case $t\ge s+2$ by
Lemma~\ref{ext:lem:wire-separation}.  In the second case, $s$ has a wire to the
first block of the clause triple, which precedes $t$ by two state blocks, so
again $t\ge s+2$.  Lemma~\ref{ext:lem:envelope} therefore places every such
constraint copy below the window of $t$.  Thus only the copy of $J$ remains,
and the occurrence uses both of its entries.
\end{proof}

\begin{lemma}[Occurrences of the overlap and boundary windows]
\label{ext:lem:boundary-occurrences}
Let $\pi$ be a configuration whose inserted maxima increase from left to right.
In every occurrence of $U_D$, the copies of $D,N,S,N$ selected by the
occurrence are, respectively, the initial copies of $D,N$, the $S$-copy in the
first state block, and its first following $N$-copy.  In every occurrence of $U$, the copies of $N,N,S,N$ selected by the occurrence are, respectively, the
final two $N$-copies of one state window, the $S$-copy in the next state
block, and its first following $N$-copy.  Every
occurrence of $U_E$ consists of the final two $N$-copies of the last state
window followed by the final copies $E,N$.
\end{lemma}

\begin{proof}
For $U_D=\vartheta[D,N,F,N]$, Lemma~\ref{ext:lem:isolation} forces the selected
copies of $D,N,S,N$ to be copies placed by the construction.  The copy of $D$
is unique and is the first block of the root.
Lemma~\ref{ext:lem:osc-window} therefore
forces the four positions of $\omega_{4r+4}$ corresponding to these copies to
be $1,2,3,4$, which are the stated four copies.

For $U=\vartheta[N,N,F,N]$, the selected copy of $S$ lies inside the third
inflated block and is a copy placed by the construction.  It lies at position
$4c+3$ for some state block $c$.
The four positions of $\omega_{4r+4}$ corresponding to these copies are
consecutive and begin at an odd position, so they are $4c+1,\dots,4c+4$.  The
case $c=0$ is impossible because its two preceding blocks are $D,N$, whereas $U$
requires $N,N$.  Thus
$c\ge1$, and the selected copies are the final two $N$-copies of state
window $c-1$, the $S$-copy in state block $c$, and its first following
$N$-copy.

For $U_E=\vartheta[N,N,E,N]$, all four inflated blocks are marker
permutations.  Lemma~\ref{ext:lem:isolation} makes them copies placed by the
construction, and the copy of $E$ is unique at position $4r+3$ of
$\omega_{4r+4}$.  Since the $E$-copy is the third inflated block,
Lemma~\ref{ext:lem:osc-window} forces positions
$4r+1,\dots,4r+4$, which are the stated four copies.
\end{proof}

\begin{lemma}[The overlap patterns do not occur]
\label{ext:lem:overlap-patterns}
Let $\pi$ be a configuration whose inserted maxima increase from left to right.
Neither $\ins_{10}(U)$ nor $\ins_{20}(U)$ occurs in $\pi$.
\end{lemma}

\begin{proof}
Suppose one of the two patterns occurs.  Deleting its largest entry leaves an
occurrence of $U$, whose marker copies identify the designated
overlap window immediately before some state block $t\ge1$, by
Lemma~\ref{ext:lem:boundary-occurrences}.

The entry playing the largest value of the forbidden pattern lies in gap $10$
or $20$ of this occurrence.  Gap $10$ contains no root entry or constraint
copy.  Gap $20$ may contain incoming copies placed immediately before $F(t)$.
The four positions of $\omega_{4r+4}$ corresponding to the occurrence are
$4t+1,\dots,4t+4$, whose defining-sequence values have minimum $4t+1$.  Any
incoming constraint copy has a source $s\le t-2$
by Lemma~\ref{ext:lem:wire-separation}.  Its value band lies inside the value
interval of $F(s)$, whose defining-sequence value is $4s+6\le4t-2$.  Hence
every such copy lies below all four blocks of the overlap window and cannot
supply the largest entry.  Neither gap is a site, so it contains no inserted
maximum.  Thus neither gap can contain the largest entry required by the
forbidden pattern, contradicting the assumed occurrence.
\end{proof}

Lemmas~\ref{ext:lem:window-occurrences},
\ref{ext:lem:boundary-occurrences},
and~\ref{ext:lem:overlap-patterns} determine the state windows and marker copies
that can occur in $A,U,U_D,U_E$, and exclude the two overlap patterns.  We can
now determine the active gaps and maximum extension depth.

\subsubsection{Open gaps and maximum depth in the base permutation}
\label{ext:sec:state-analysis}

This section analyzes the base permutation $\rho^{\star}_r$ under $\Bstate$.

A \emph{site configuration} is obtained from $\rho^{\star}_r$ by choosing at
most one of $P_0,P_1$ in each state block and inserting one maximum into each
chosen site in left-to-right order.

\begin{lemma}[Site configurations avoid $\Bstate$]
\label{ext:lem:compatible-avoid}
Every site configuration avoids $\Bstate$.  In particular,
$\rho^{\star}_r\in\Av(\Bstate)$.
\end{lemma}

\begin{proof}
We check each family in Table~\ref{ext:tab:categories}.  Throughout, $\pi$ is
the configuration.  Its inserted maxima lie at sites and increase from left
to right by definition.  Hence Lemmas~\ref{ext:lem:isolation},
\ref{ext:lem:window-occurrences}, \ref{ext:lem:boundary-occurrences},
and~\ref{ext:lem:overlap-patterns} apply.

\emph{$\mathcal B_{\mathrm{core}}$.}
By Lemma~\ref{ext:lem:isolation}, an occurrence of $\ins_g(T)$ would need a
marker copy of $T$ together with an extra entry above it, positioned strictly
inside it.  No inserted maximum and no other entry lies inside a marker copy.

\emph{$\mathcal B_{\mathrm{order}}$.}  An occurrence of
$4312[1,1,N,E]$ needs copies of $N$ and $E$ placed by the
construction together with two entries  above $E$, to their left, in decreasing
order.  Such entries  would be inserted maxima, which increase from left to right.

\emph{$\mathcal B_{\mathrm{gap}}$.}  By
Lemma~\ref{ext:lem:window-occurrences}, an occurrence of $\ins_g(A)$,
$g\in\{12,22\}$, forces one state window plus an entry above it in the
corresponding root gap.  Those gaps are not sites, so no such entry exists.

\emph{$\mathcal B_{\mathrm{cap}}$ and $\mathcal B_{\mathrm{conf}}$.}
These need two maxima in the same site or in both sites of one state block.
This is excluded by the definition of a site configuration.

\emph{$\mathcal B_{\mathrm{overlap}}$.}  These patterns are excluded by
Lemma~\ref{ext:lem:overlap-patterns}.

\emph{$\mathcal B_{\mathrm{boundary}}$.}
Lemma~\ref{ext:lem:boundary-occurrences} shows that every occurrence of $U_D$
and $U_E$ uses the specified copies.  A site configuration has
no entry before $D$, in either of the two gaps of $U_D$ used by the boundary
patterns, or in any of the four relevant gaps of $U_E$.  Hence no boundary
pattern occurs.
\end{proof}

\paragraph{The active gaps are the sites}

We can now classify all gaps of the base permutation and prove that only
the two designated gaps of each state block remain active.

\begin{lemma}[The active gaps are the sites]
\label{ext:lem:gap-cover}
Let $r\ge1$.  In $\rho^{\star}_r$, every gap is inactive for $\Bstate$ except
the $2r$ sites.  Consequently, every inserted entry in an extension of
$\rho^{\star}_r$ lies at a site.
\end{lemma}

\begin{proof}
The root $\rho^{\star}_r$ avoids $\Bstate$ by
Lemma~\ref{ext:lem:compatible-avoid}.  We show that inserting a maximum into
any non-site gap creates a member of $\Bstate$.

\emph{$\mathcal B_{\mathrm{core}}$.}
The nine interior gaps of every marker copy inducing $T$ are made inactive
by $\ins_g(T)$, $1\le g\le9$.

\emph{$\mathcal B_{\mathrm{gap}}$.}
Apart from the interior gaps of marker copies, the gaps numbered within every
state window are $10,11,12,22,32,42$.  Gaps $10$ and $11$ are the sites.
Gaps $12$ and $22$ are made inactive by $\ins_{12}(A)$ and $\ins_{22}(A)$.

\emph{$\mathcal B_{\mathrm{overlap}}$.}
For a nonfinal state block $c$, the occurrence of $U$ associated with state
block $c+1$ begins after the first $22$ entries  of the window of $c$.  Its gaps
$10$ and $20$ are therefore exactly gaps $32$ and $42$ of that window.  The
patterns $\ins_{10}(U)$ and $\ins_{20}(U)$ make them inactive.

\emph{$\mathcal B_{\mathrm{boundary}}$.}
The pattern $\ins_0(D)$ makes the initial root gap inactive.  In the occurrence
of $U_D$ described by Lemma~\ref{ext:lem:boundary-occurrences}, gaps $10$ and
$20$ are, respectively, the gap after $D$ and the gap between the initial $N$
and the first state block.  In the occurrence of $U_E$ described by the same
lemma, gaps $10$ and $20$ are gaps $32$ and $42$ of the final state window,
gap $30$ is the gap after $E$, and gap $40$ is the final root gap.  The
corresponding six boundary patterns make these gaps inactive.

Every non-site gap is shown inactive in one of these cases.  Each site is active
because the configuration using only that site is a site configuration and
therefore avoids $\Bstate$ by Lemma~\ref{ext:lem:compatible-avoid}.  Finally,
deleting all but one inserted entry and standardizing shows that an
inactive root gap remains unused in every avoiding extension.  Hence every inserted entry
in an extension lies at a site.
\end{proof}

\begin{lemma}[At most one site is used per block]
\label{ext:lem:capacity}
In every extension of $\rho^{\star}_r$ in $\Av(\Bstate)$, each site receives
at most one inserted maximum, and at most one of the two sites of a state block
is used.
\end{lemma}

\begin{proof}
By Lemma~\ref{ext:lem:ordering}, the inserted maxima increase from left to
right. Suppose two inserted maxima have the same root gap,
corresponding to site $g$ of a state block.  Together with its window they form an occurrence of
$\ins^{\nearrow2}_g(A)$, contradicting $\Bstate$-avoidance.  If both sites of
one state block are used, the two maxima together with its window form an
occurrence of $\ins^{\nearrow}_{10,11}(A)$, again contradicting
$\Bstate$-avoidance.
\end{proof}

\begin{theorem}[Depth of the base permutation]
\label{ext:thm:depth}
For every $r\ge1$, $d_{\Bstate}\bigl(\rho^{\star}_r\bigr)=r$.
Moreover, for every $(b_0,\dots,b_{r-1})\in\{0,1\}^r$, there is an extension
of depth $r$ in which state block $c$ is assigned bit $b_c$ for every
$c\in\{0,\dots,r-1\}$, and every extension of depth $r$ arises from exactly
one such tuple.
\end{theorem}

\begin{proof}
\emph{Upper bound.}  By Lemma~\ref{ext:lem:gap-cover}, every inserted maximum
lies at a site.  By Lemma~\ref{ext:lem:capacity}, at most one site is used in
each state block.  Thus every extension has depth at most $r$.

\emph{Lower bound and Boolean states.}  Choose a bit $b_c$ for each
block $c$ and insert a maximum at site $P_{b_c}$, processing all $r$ chosen
sites from left to right.  The resulting site configuration avoids $\Bstate$
by Lemma~\ref{ext:lem:compatible-avoid} and is therefore an extension of depth
$r$.  Conversely, an extension of depth $r$ must use exactly one site in every
state block, and it is $P_b$ for a unique $b\in\{0,1\}$, so the block is assigned bit $b$. By Lemma~\ref{ext:lem:ordering}, the inserted entries occur in increasing order from left to right. 
\end{proof}

\subsubsection{Constraint copies preserve \texorpdfstring{$\Bstate$}{Bstate}-avoidance}
\label{ext:sec:preservation}

Theorem~\ref{ext:thm:depth} gives two independent choices in each state block.
To obtain $\rho_\Phi$, we add the constraint copies.  We now prove that these copies
do not create active gaps or occurrences of members of $\Bstate$.

\begin{lemma}[Constraint copies create no new active gaps]
\label{ext:lem:aux-gaps}
Every inserted entry in an extension of $\rho_\Phi$ in $\Av(\Bstate)$ lies at
a site.
\end{lemma}

\begin{proof}
Deleting the entries of all constraint copies and standardizing maps
$\rho_\Phi$ to $\rho^{\star}_r$.
By~\ref{ext:D1}--\ref{ext:D3}, this deletion preserves every site and maps
every non-site gap to a non-site gap of $\rho^{\star}_r$.
Suppose an extension $\sigma$ contains an inserted entry $x$ outside the
sites.  Delete all entries of constraint copies and all inserted entries except
$x$, then standardize. The remaining permutation is $\ins_g(\rho^{\star}_r)$ for a non-site
gap $g$, so it does not avoid $\Bstate$ by
Lemma~\ref{ext:lem:gap-cover}. However, it is a pattern of $\sigma$ and hence avoids
$\Bstate$, a contradiction. Thus every inserted entry
lies at a site.
\end{proof}

An \emph{assignment configuration} of $\rho_\Phi$ is obtained by choosing
$b_c\in\{0,1\}$ for every state block $c$ and inserting a maximum at the site
$P_{b_c}$, processing all chosen sites from left to right.  It contains exactly
$r$ inserted maxima.

\begin{theorem}[Constraint copies preserve $\Bstate$-avoidance]
\label{ext:thm:preservation}
Every assignment configuration of $\rho_\Phi$ avoids $\Bstate$.
\end{theorem}

\begin{proof}
Let $\pi$ be an assignment configuration, and let $\bar\pi$ be obtained by
deleting the entries of its constraint copies and standardizing.
Then $\bar\pi$ is a site configuration
and avoids $\Bstate$ by Lemma~\ref{ext:lem:compatible-avoid}.  Hence any
occurrence in $\pi$ of a member of $\Bstate$ must use an entry of a constraint copy.
By definition of an assignment configuration, the inserted maxima of $\pi$
increase from left to right.  Hence
Lemmas~\ref{ext:lem:isolation}--\ref{ext:lem:overlap-patterns} apply to $\pi$.

By~\ref{ext:D1}--\ref{ext:D3}, no constraint copy lies inside a marker copy or
at a site.  Thus constraint copies cannot create occurrences from
$\mathcal B_{\mathrm{core}}$, $\mathcal B_{\mathrm{cap}}$, or
$\mathcal B_{\mathrm{conf}}$.  They also cannot create $4312[1,1,N,E]$.  Every
constraint-copy entry lies below the final copy of $E$, so the two entries  above
$E$ would have to be inserted maxima in decreasing left-to-right order.

For $\mathcal B_{\mathrm{gap}}$, Lemma~\ref{ext:lem:window-occurrences}
forces the required window.  Among its two relevant gaps, only gap $12$ can
contain constraint-copy entries.  Those entries  lie below the window by
Lemma~\ref{ext:lem:envelope}, while the extra entry of the forbidden pattern
must lie above it.  The overlap patterns are excluded directly by
Lemma~\ref{ext:lem:overlap-patterns}.

For the left boundary, Lemma~\ref{ext:lem:boundary-occurrences} places the
selected copies of $D,N$ at the start of the root and the selected copy of
$S$ in the first state block.  The first state block is in level $0$ and has
no incoming wire, while every constraint copy whose source is that block is
positioned at a later target.  Hence no constraint copy lies after $D$ or
between the initial $N$ and the first state block.  For the right boundary,
the same lemma places the occurrence of $U_E$ on the final two $N$-copies of
the last state window and the final copies $E,N$.  Constraint copies targeting
the final state block lie immediately before or after its copy of $F$, hence
before its first following $N$-copy, while the four relevant gaps of $U_E$
begin after its second following $N$-copy.  Thus no boundary pattern occurs.
\end{proof}

\begin{corollary}[Depth of $\rho_\Phi$]
\label{ext:cor:state-depth}
$\rho_\Phi\in\Av(\Bstate)$ and
$d_{\Bstate}(\rho_\Phi)=r$.
\end{corollary}

\begin{proof}
An assignment configuration exists, contains $r$ inserted maxima, and avoids
$\Bstate$ by Theorem~\ref{ext:thm:preservation}.  Hence
$d_{\Bstate}(\rho_\Phi)\ge r$.  Since $\rho_\Phi$ is obtained from this
configuration by deleting the inserted maxima, hereditary avoidance also
gives $\rho_\Phi\in\Av(\Bstate)$.

For any extension of $\rho_\Phi$ in $\Av(\Bstate)$, delete the
constraint-copy entries and standardize. This gives an extension of
$\rho^{\star}_r$ at the same depth, so
Theorem~\ref{ext:thm:depth} bounds that depth by $r$.
Thus $d_{\Bstate}(\rho_\Phi)=r$.
\end{proof}

Thus the extension depth of $\rho_\Phi$ in $\Av(\Bstate)$ is $r$,
and every extension of depth $r$ chooses one of the two Boolean states in
each block, exactly as for the base permutation.  It remains to characterize
occurrences of the violation patterns, beginning with $\Pneq{01}$ and
$\Pneq{10}$.

\subsubsection{Occurrences of the violation patterns}
\label{ext:sec:violation-occurrences}

The violation patterns were defined in
Section~\ref{ext:sec:violation-patterns}.  We now prove that every occurrence
of a violation pattern comes from a wire or clause triple.

\paragraph{Equality along wires}

\begin{figure}[ht]
\centering
\begin{tikzpicture}[x=0.68cm,y=0.34cm,every node/.style={font=\scriptsize}]
  \draw[-{Latex[length=1.5mm]},black!50] (-0.8,-0.8)--(14.2,-0.8)
    node[anchor=west,font=\tiny] {block position};
  \draw[-{Latex[length=1.5mm]},black!50] (-0.8,-0.8)--(-0.8,19.5)
    node[anchor=south,font=\tiny,rotate=90] {block value};

  \foreach \x/\top in {5.8/3.7,7/12.5,8.8/4.8}
    {\draw[black!25,densely dashed,line width=0.45pt]
       (\x,-0.72)--(\x,\top);}

  \node[draw=blue!65!black,fill=blue!18] (eqsource)
    at (0,4.5) {$F(s)$};
  \foreach \x/\y in {1.2/0.5,2.4/6.5,3.6/1.5}
    {\node[draw=blue!65!black,fill=blue!7] at (\x,\y) {$N$};}
  \node[draw,fill=orange!30] (qlcopy) at (5.8,3.7) {$Q_L(s)$};
  \node[draw=violet!75!black,fill=violet!16] at (7,12.5) {$F(t)$};
  \node[draw,fill=orange!30] (qrcopy) at (8.8,4.8) {$Q_R(s)$};
  \coordinate (eqbandlow) at
    ($(eqsource.south east)!0.20!(eqsource.north east)$);
  \coordinate (eqbandhigh) at
    ($(eqsource.south east)!0.80!(eqsource.north east)$);
  \draw[densely dotted,orange!80!black,line width=0.7pt]
    (eqbandlow)--(qlcopy.west |- eqbandlow);
  \draw[densely dotted,orange!80!black,line width=0.7pt]
    (eqbandhigh)--(qrcopy.west |- eqbandhigh);
  \foreach \x/\y in {10.4/8.5,11.6/14.5,12.8/9.5}
    {\node[draw=violet!75!black,fill=violet!7] at (\x,\y) {$N$};}
  \draw[draw=blue!65!black,fill=blue!12]
    (1.0,-1.95) rectangle (1.40,-1.50);
  \node[anchor=west,blue!60!black,font=\scriptsize] at (1.55,-1.725)
    {window of $s$};
  \draw[draw=violet!75!black,fill=violet!12]
    (7.5,-1.95) rectangle (7.90,-1.50);
  \node[anchor=west,violet!70!black,font=\scriptsize] at (8.05,-1.725)
    {window of $t$};
  \foreach \x/\y/\lab in {0.2/16.5/{P_0},7.2/18/{P_1}}
    {\draw[blue!70!black,line width=1.1pt]
       (\x-0.22,\y-0.22)--(\x+0.22,\y+0.22)
       (\x-0.22,\y+0.22)--(\x+0.22,\y-0.22);
     \node[anchor=south,blue!55!black] at (\x,\y+0.42) {$\lab$};}
\end{tikzpicture}
\caption{Block-level plot of
$\Pneq{01}=\ins^{\nearrow}_{10,63}(\Aedge)$.  The orange copies of $Q_L$ and $Q_R$ lie in the
value band of the source but bracket the target state block in position.
Matching outline colors identify the four blocks of the source and target
windows.  In particular, $Q_R(s)$ is not part of the window of $t$.  The two
inserted maxima occupy site $P_0$ of $F(s)$ and site $P_1$ of $F(t)$, so this pattern detects
the forbidden unequal assignment $(b_s,b_t)=(0,1)$.  The pattern
$\Pneq{10}$ is obtained by using the other site in each state block.}
\label{ext:fig:Pneq}
\end{figure}
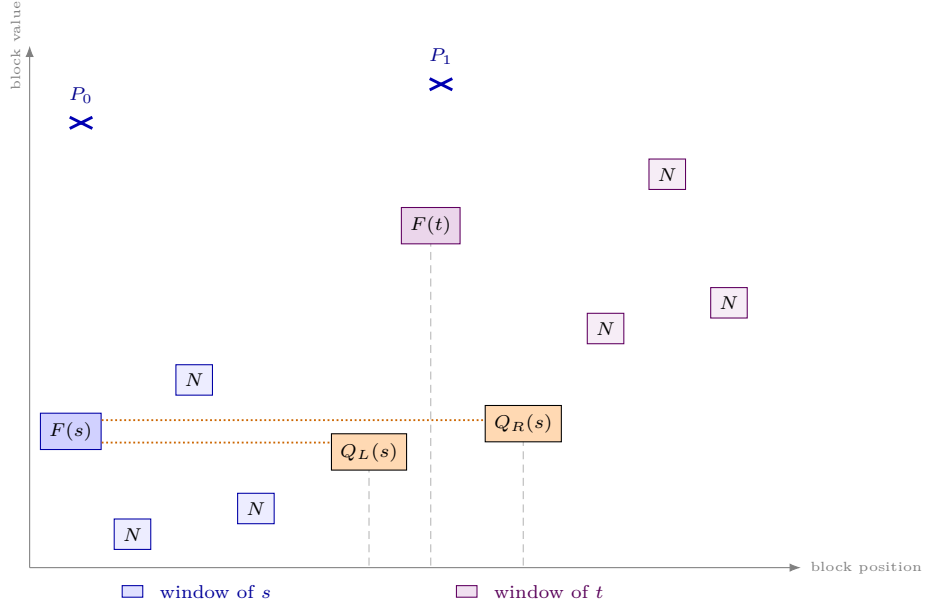

Throughout this subsection, $\pi$ is an assignment configuration of
$\rho_\Phi$.  For each state block $c$, the notation $b_c$ denotes the unique
bit assigned to $c$ in $\pi$.

\begin{lemma}[Occurrences of $\Aedge$]
\label{ext:lem:Aeq-faithful}
For every occurrence of $\Aedge$ in $\pi$, there is a wire
$s\to t\in\mathcal W_\Phi$ such that the occurrence consists exactly of the
windows of $s$ and $t$ together with the copies $Q_L(s)$ and $Q_R(s)$.
\end{lemma}

\begin{proof}
Let $O$ be an occurrence of $\Aedge$.  Deleting the entries  in its $Q_L$- and
$Q_R$-blocks leaves two occurrences of $A$.  By
Lemma~\ref{ext:lem:window-occurrences}, they are the windows of state blocks
$s$ and $t$, in that order.

By Lemma~\ref{ext:lem:isolation}, the deleted entries  form copies of $Q_L$ and
$Q_R$ placed by the construction.  Their values place both copies in the
value band of $s$.
Rules~\ref{ext:D1} and~\ref{ext:D3} therefore identify them as $Q_L(s)$ and
$Q_R(s)$ for a wire with source $s$.  The state block of the second window is
$F(t)$ and lies positionally between these copies.  By
Lemma~\ref{ext:lem:bracket}, it is the target of that wire.  Hence $s\to t$ is
a wire, and $O$ has the required form.
\end{proof}

\begin{lemma}[$\Pneq{01}$ and $\Pneq{10}$ detect unequal endpoints]
\label{ext:lem:edge-detection}
For each $(a,b)\in\{(0,1),(1,0)\}$,
the occurrences of $\Pneq{ab}$ in $\pi$ are exactly those obtained from wires
$s\to t$ satisfying $(b_s,b_t)=(a,b)$ by adding the inserted maxima at
site $P_a$ of $F(s)$ and site $P_b$ of $F(t)$ to the associated
$\Aedge$-occurrence.
\end{lemma}

\begin{proof}
Fix $(a,b)\in\{(0,1),(1,0)\}$.

Suppose first that $(b_s,b_t)=(a,b)$ for a wire $s\to t$.  The
corresponding $\Aedge$-occurrence is present in $\pi$, and site $P_a$ of $F(s)$ and site $P_b$ of $F(t)$ contain inserted maxima.  The first maximum lies to the left of
the second, so Lemma~\ref{ext:lem:ordering} makes their values increase from
left to right.  By the definition of $\Pneq{ab}$, these entries  form an
occurrence of $\Pneq{ab}$.

Conversely, fix an occurrence $O$ of $\Pneq{ab}$ and remove its two
largest entries.  The remaining entries  form an occurrence of $\Aedge$, so
Lemma~\ref{ext:lem:Aeq-faithful} associates them with a wire $s\to t$.

By the definition of $\Pneq{ab}$, the deleted entries lie at site $P_a$ of $F(s)$ and
site $P_b$ of $F(t)$.  Since $\pi$ is an assignment configuration, they are
inserted maxima.  Thus these two sites are used, giving
$(b_s,b_t)=(a,b)$.
\end{proof}

\begin{corollary}[Avoidance enforces equality]
\label{ext:cor:edge-avoidance}
$\pi$ avoids $\{\Pneq{01},\Pneq{10}\}$ if and only if $b_s=b_t$ for every
wire $s\to t\in\mathcal W_\Phi$.
\end{corollary}

\begin{proof}
The two forbidden bit pairs are exactly the two unequal pairs in $\{0,1\}^2$.
\end{proof}

\paragraph{Not-all-equal clauses}

\begin{figure}[ht]
\centering
\begin{tikzpicture}[x=0.46cm,y=0.27cm,every node/.style={font=\scriptsize}]
  \draw[-{Latex[length=1.5mm]},black!50] (-0.8,-0.8)--(24.7,-0.8)
    node[anchor=west,font=\tiny] {block position};
  \draw[-{Latex[length=1.5mm]},black!50] (-0.8,-0.8)--(-0.8,31)
    node[anchor=south,font=\tiny,rotate=90] {block value};

  \foreach \x/\top in {5.8/3.4,7/12,12.4/16,17.8/20,19.6/5.2}
    {\draw[black!25,densely dashed,line width=0.45pt]
       (\x,-0.72)--(\x,\top);}

  \node[draw=blue!65!black,fill=blue!18,inner sep=2pt,font=\tiny]
    (naesource) at (0,4.5) {$F(s_0)$};
  \foreach \x/\y in {1.2/0.5,2.4/6.5,3.6/1.5}
    {\node[draw=blue!65!black,fill=blue!7] at (\x,\y) {$N$};}
  \node[draw,fill=orange!30] (clcopy) at (5.8,3.4) {$C_L$};

  \node[draw=violet!75!black,fill=violet!16,inner sep=2pt,font=\tiny]
    at (7,12) {$F(t_0)$};
  \foreach \x/\y in {8.2/9,9.4/14,10.6/11}
    {\node[draw=violet!75!black,fill=violet!7] at (\x,\y) {$N$};}
  \node[draw=green!50!black,fill=green!13,inner sep=2pt,font=\tiny]
    at (12.4,16) {$F(t_1)$};
  \foreach \x/\y in {13.6/13,14.8/18,16.0/15}
    {\node[draw=green!50!black,fill=green!6] at (\x,\y) {$N$};}
  \node[draw=red!70!black,fill=red!13,inner sep=2pt,font=\tiny]
    at (17.8,20) {$F(t_2)$};
  \node[draw,fill=orange!30] (crcopy) at (19.6,5.2) {$C_R$};
  \coordinate (naebandlow) at
    ($(naesource.south east)!0.20!(naesource.north east)$);
  \coordinate (naebandhigh) at
    ($(naesource.south east)!0.80!(naesource.north east)$);
  \draw[densely dotted,orange!80!black,line width=0.7pt]
    (naebandlow)--(clcopy.west |- naebandlow);
  \draw[densely dotted,orange!80!black,line width=0.7pt]
    (naebandhigh)--(crcopy.west |- naebandhigh);
  \foreach \x/\y in {20.8/17,22.0/22,23.2/19}
    {\node[draw=red!70!black,fill=red!6] at (\x,\y) {$N$};}
  \draw[draw=blue!65!black,fill=blue!12] (0.3,-2.15) rectangle (0.70,-1.70);
  \node[anchor=west,blue!60!black,font=\scriptsize] at (0.85,-1.925)
    {window of $s_0$};
  \draw[draw=violet!75!black,fill=violet!12]
    (6.0,-2.15) rectangle (6.40,-1.70);
  \node[anchor=west,violet!70!black,font=\scriptsize] at (6.55,-1.925)
    {window of $t_0$};
  \draw[draw=green!50!black,fill=green!10]
    (11.4,-2.15) rectangle (11.80,-1.70);
  \node[anchor=west,green!40!black,font=\scriptsize] at (11.95,-1.925)
    {window of $t_1$};
  \draw[draw=red!70!black,fill=red!10]
    (16.8,-2.15) rectangle (17.20,-1.70);
  \node[anchor=west,red!65!black,font=\scriptsize] at (17.35,-1.925)
    {window of $t_2$};

  \foreach \x/\y/\lab in {7.2/26/{P_0},12.6/27.5/{P_0},18.0/29/{P_0}}
    {\draw[blue!70!black,line width=1.1pt]
       (\x-0.22,\y-0.32)--(\x+0.22,\y+0.32)
       (\x-0.22,\y+0.32)--(\x+0.22,\y-0.32);
     \node[anchor=south,blue!55!black] at (\x,\y+0.42) {$\lab$};}
\end{tikzpicture}
\caption{Block-level plot of $\Pmono{000}$.  The four blocks of each state
window have relative order $3142$, and matching outline colors identify
which four blocks form each window.  The copies of $C_L$ and $C_R$ lie in the value
band of the source but occupy positions bracketing the three target state
blocks. The three inserted maxima use the
$P_0$ sites, detecting the forbidden assignment $000$.
$\Pmono{111}$ instead uses the three $P_1$ sites.}
\label{ext:fig:Pmono}
\end{figure}
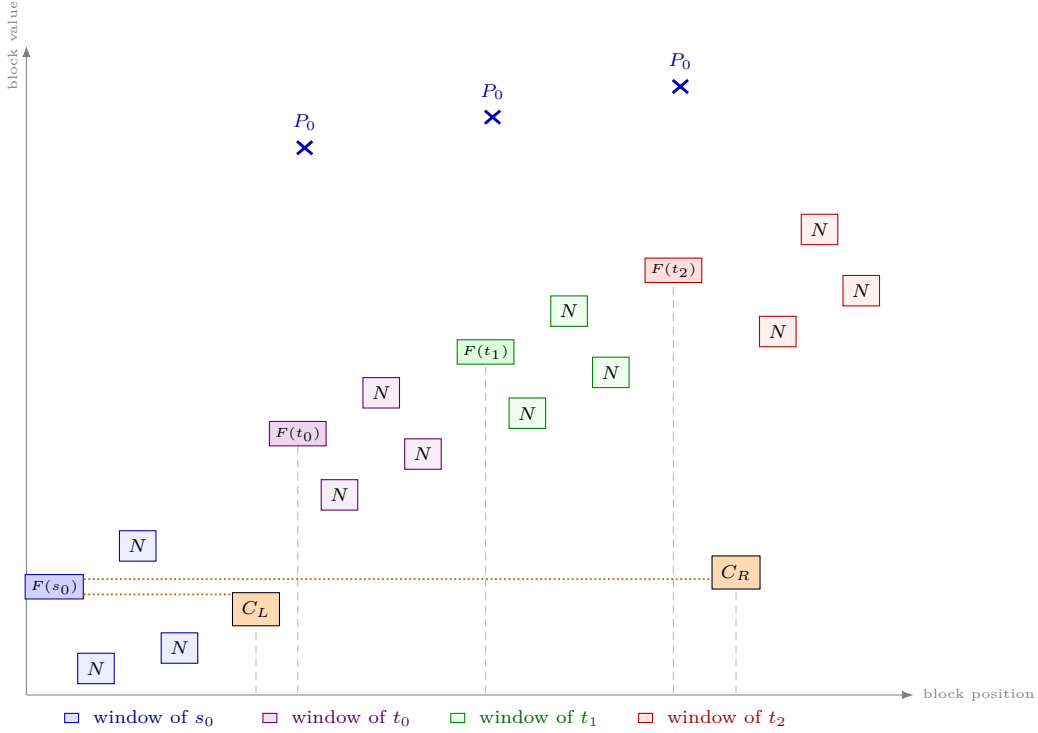

For a clause triple $\mathbf t=(t_0,t_1,t_2)\in\mathcal C_\Phi$, put
$b(\mathbf t)=(b_{t_0},b_{t_1},b_{t_2})$.

\begin{lemma}[Occurrences of $\Anae$]
\label{ext:lem:Anae-faithful}
For every occurrence of $\Anae$ in $\pi$, there is a clause triple
$\mathbf t=(t_0,t_1,t_2)\in\mathcal C_\Phi$ such that the occurrence consists
exactly
of the windows of $s_0,t_0,t_1,t_2$ together with the copies $C_L(s_0)$ and
$C_R(s_0)$, where $s_0$ is the source of the incoming wire to $t_0$.
\end{lemma}

\begin{proof}
Let $O$ be an occurrence of $\Anae$.  Deleting the entries  in its $C_L$- and
$C_R$-blocks leaves four occurrences of $A$.  By
Lemma~\ref{ext:lem:window-occurrences}, they are the windows of state blocks
$s_0,t_0,t_1,t_2$ in the order prescribed by $\Anae$.

By Lemma~\ref{ext:lem:isolation}, the deleted entries form copies of $C_L$ and $C_R$
placed by the construction.  Their values place both copies in the value band
of $s_0$.
Rules~\ref{ext:D2} and~\ref{ext:D3} therefore identify them as $C_L(s_0)$ and
$C_R(s_0)$ for a clause triple whose incoming wire to its first state block
has source $s_0$.  By
Lemma~\ref{ext:lem:bracket}, the three state blocks positionally between these
copies are exactly the targets of that triple, in order.  Hence the triple is
$\mathbf t=(t_0,t_1,t_2)$, and $O$ consists exactly of the copies and windows
stated in the lemma.
\end{proof}

\begin{lemma}[$\Pmono{000}$ and $\Pmono{111}$ detect constant triples]
\label{ext:lem:nae-detection}
The occurrences of $\Pmono{000}$ in $\pi$ are exactly those obtained from
clause triples $\mathbf t=(t_0,t_1,t_2)$ satisfying
$b(\mathbf t)=(0,0,0)$ by adding the
inserted maxima at site $P_0$ of each $F(t_i)$, $i=0,1,2$ to the associated
$\Anae$-occurrence.  The occurrences of $\Pmono{111}$ are described
analogously, with $1$ in place of $0$.
\end{lemma}

\begin{proof}
Fix $a\in\{0,1\}$, and let $P_a^{\mathrm{viol}}$ denote $\Pmono{000}$ when
$a=0$ and $\Pmono{111}$ when $a=1$.

Suppose first that $b(\mathbf t)=(a,a,a)$ for a clause triple
$\mathbf t=(t_0,t_1,t_2)$.  The associated $\Anae$-occurrence is present in $\pi$,
and site $P_a$ of each $F(t_i)$, $i=0,1,2$, contains an inserted maximum.
These maxima increase from left to right by Lemma~\ref{ext:lem:ordering}, so
the definition of $P_a^{\mathrm{viol}}$ gives an occurrence of
$P_a^{\mathrm{viol}}$.

Conversely, fix an occurrence $O$ of $P_a^{\mathrm{viol}}$ and remove its
three largest entries.  The remaining entries  form an occurrence of $\Anae$, so
Lemma~\ref{ext:lem:Anae-faithful} associates them with a clause triple
$\mathbf t=(t_0,t_1,t_2)$.

By the definition of $P_a^{\mathrm{viol}}$, the deleted entries lie at
site $P_a$ of each $F(t_i)$, $i=0,1,2$.  Since $\pi$ is an assignment
configuration, they are inserted maxima.  Thus these three sites are used,
giving $b(\mathbf t)=(a,a,a)$.
\end{proof}

\begin{corollary}[Avoidance enforces NAE]
\label{ext:cor:nae-avoidance}
$\pi$ avoids $\{\Pmono{000},\Pmono{111}\}$ if and only if
$b(\mathbf t)$ is not constant for every clause triple
$\mathbf t\in\mathcal C_\Phi$.
\end{corollary}

\begin{proof}
By Lemma~\ref{ext:lem:nae-detection}, the two forbidden patterns occur exactly
on the two constant triples $000$ and $111$.
\end{proof}

Later levels create no additional $\Anae$-occurrences.
Lemma~\ref{ext:lem:Anae-faithful} applies to the whole permutation $\pi$, not
only to the clause level.  Thus every occurrence of $\Anae$ uses the source
window and the windows of the three state blocks in one clause triple.
By~\ref{ext:D2},
copies of $C_L$ and $C_R$ occur only in the value bands of the sources of
the incoming wires to the first blocks of clause triples, and
Lemma~\ref{ext:lem:bracket} forces the corresponding
three state blocks.  Later levels add copies of $Q_L$ and $Q_R$ and state blocks but no
copies of $C_L$ or $C_R$, so they create no further occurrence of $\Anae$.

We have now proved that $\Bstate$ gives two states per block, that adding
constraint copies preserves $\Bstate$-avoidance, and that the violation
patterns occur exactly on unequal wires and constant clause triples.

\FloatBarrier

\subsection{Correctness of the formula instance}
\label{ext:sec:formula-correctness}

We now combine Theorem~\ref{ext:thm:depth},
Theorem~\ref{ext:thm:preservation}, and
Corollaries~\ref{ext:cor:edge-avoidance} and~\ref{ext:cor:nae-avoidance} to
prove that the constructed instance has the required depth exactly when the
formula is satisfiable.

\begin{lemma}[The permutation $\rho_\Phi$ avoids $\Bfinal$]
\label{ext:lem:root-avoidance}
$\rho_\Phi$ lies in $\Av(\Bfinal)$.
\end{lemma}

\begin{proof}
Choose any assignment configuration $\pi$ of $\rho_\Phi$.  By
Theorem~\ref{ext:thm:preservation}, $\pi$ avoids $\Bstate$.  Since $\rho_\Phi$
is obtained from $\pi$ by deleting the inserted maxima, hereditary avoidance
gives $\rho_\Phi\in\Av(\Bstate)$.

Suppose that $\rho_\Phi$ contained a violation pattern.  The same root entries
would form an occurrence in $\pi$.  Lemmas~\ref{ext:lem:edge-detection}
and~\ref{ext:lem:nae-detection}, however, show that every such occurrence in
$\pi$
uses two or three inserted maxima.  This is impossible for an occurrence
formed entirely from root entries.  Thus $\rho_\Phi$ also avoids
$\mathcal B_{\mathrm{viol}}$, and Definition~\ref{ext:def:bfinal} gives
$\rho_\Phi\in\Av(\Bfinal)$.
\end{proof}

\begin{theorem}[Correctness of the formula instance]
\label{ext:thm:encoding}
The permutation $\rho_\Phi$ lies in $\Av(\Bfinal)$, and
\[
  d_{\Bfinal}(\rho_\Phi)\le r.
\]
Equality holds if and only if $\Phi$ is NAE-satisfiable.
\end{theorem}

\begin{proof}
Lemma~\ref{ext:lem:root-avoidance} gives
$\rho_\Phi\in\Av(\Bfinal)$.  Since $\Bstate\subseteq\Bfinal$, every
$\Bfinal$-avoiding extension is also a $\Bstate$-avoiding extension.  Hence
Corollary~\ref{ext:cor:state-depth} gives
\[
  d_{\Bfinal}(\rho_\Phi)
  \le d_{\Bstate}(\rho_\Phi)
  = r.
\]

Suppose first that $\Phi$ is NAE-satisfiable.  By
Lemma~\ref{ext:lem:equivalence}, choose a bit $b_c\in\{0,1\}$ for each state
block $c$ such that the chosen bits are equal along every wire and nonconstant
on every clause triple.  Insert a maximum at $P_{b_c}$ in each state block,
processing the selected sites from left to right.  The resulting assignment
configuration avoids $\Bstate$
by Theorem~\ref{ext:thm:preservation}, $\Pneq{01}$ and $\Pneq{10}$ by
Corollary~\ref{ext:cor:edge-avoidance}, and $\Pmono{000}$ and $\Pmono{111}$
by Corollary~\ref{ext:cor:nae-avoidance}.  It therefore avoids $\Bfinal$.
It contains $r$ inserted entries and is therefore an extension of
$\rho_\Phi$ at depth $r$.

Conversely, let $\pi$ be a $\Bfinal$-avoiding extension of $\rho_\Phi$ at
depth $r$.
Lemma~\ref{ext:lem:aux-gaps} places every inserted entry at a site. Deleting the constraint-copy entries and standardizing gives an
extension of $\rho^{\star}_r$ at the same depth.  By Theorem~\ref{ext:thm:depth}, it uses the site $P_{b_c}$ for a unique
bit $b_c$ in every state block.  Therefore $\pi$ is
the corresponding assignment configuration of $\rho_\Phi$.

Since $\pi$ avoids the violation patterns,
Corollaries~\ref{ext:cor:edge-avoidance} and~\ref{ext:cor:nae-avoidance} show
that these bits are equal along every wire and nonconstant on every clause
triple.  Lemma~\ref{ext:lem:equivalence} now gives an NAE-satisfying assignment
of $\Phi$.
\end{proof}

The same theorem also gives the promised unbounded-pocket-depth result.
There are NAE-satisfiable instances with arbitrarily many disjoint clauses,
so the corresponding values of $r$ are unbounded.  For each such instance,
Theorem~\ref{ext:thm:encoding} gives a mortal permutation $\rho_\Phi$ with
$d_{\Bfinal}(\rho_\Phi)=r$.  Every mortal permutation is contained in a
pocket whose depth is at least $d_{\Bfinal}(\rho_\Phi)=r$.  Hence
$\Av(\Bfinal)$ has unbounded pocket depth.

\subsection{\NPhardness{} and \NPcompleteness{}}
\label{ext:sec:correctness}

Theorem~\ref{ext:thm:encoding} establishes the correctness of the construction.
The following theorem gives its complexity consequence.

\begin{theorem}[\NPhardness{}]
\label{ext:thm:hard}
The unary-target problem $\ExtendMax_{\Bfinal}^{\mathrm{un}}$ is
\NPhard{}, and the same holds for the binary-target problem
$\ExtendMax_{\Bfinal}^{\mathrm{bin}}$.
\end{theorem}

\begin{proof}
The construction is polynomial by Equation~\eqref{ext:eq:size} and correct by
Theorem~\ref{ext:thm:encoding}.  The source problem is \NPcomplete{} by
Theorem~\ref{ext:thm:source-hard}.  Since $r$ is polynomial in $|\Phi|$, its
unary encoding has polynomial length.  The construction therefore proves that
$\ExtendMax_{\Bfinal}^{\mathrm{un}}$ is \NPhard{}.  Binary-target
hardness follows by encoding the same polynomially bounded target
$r$ in binary.
\end{proof}

\begin{proof}[Proof of Theorem~\ref{thm:main-hardness}]
We have already specified the $55$ permutations of size at most 191 in $\Bfinal$. 
Theorem~\ref{ext:thm:hard} gives \NPhardness{}.  Membership in NP follows from
Theorem~\ref{thm:extendmax-in-np} applied
to $\Bfinal$ of Definition~\ref{ext:def:bfinal}. 
\end{proof}

The equivalence from
Proposition~\ref{prop:existence-enumeration-equivalence} now gives an enumeration
consequence for the same class.

\begin{corollary}
\label{ext:cor:rooted-enumeration-no-delayp}
Unless $\mathrm P=\mathrm{NP}$,
$\EnumExt_{\Bfinal}\notin\mathrm{DelayP}$.
\end{corollary}

\begin{proof}
Theorem~\ref{thm:main-hardness} makes the unary existence problem
$\ExtendMax_{\Bfinal}^{\mathrm{un}}$ \NPcomplete{}.  If
$\EnumExt_{\Bfinal}$ had polynomial
delay, Proposition~\ref{prop:existence-enumeration-equivalence} would put that existence
problem in $\mathrm P$.
\end{proof}

This does not show that $\EnumAvoid_{\Bfinal}\notin\mathrm{DelayP}$.
The problem $\EnumExt_{\Bfinal}$ allows an arbitrary starting permutation
$\rho$, whereas $\EnumAvoid_{\Bfinal}$ fixes the start to $\varepsilon$.
More generally, it remains open whether there exists any fixed finite set
$B$ for which $\EnumAvoid_B\notin\mathrm{DelayP}$.

\subsection[\NPcompleteness{} of the forbidden permutation CSP]{%
  \NPcompleteness{} of the forbidden permutation CSP}
\label{ext:sec:hanika-fpcsp}

The forbidden permutation CSP introduced by Hanika allows both linear orders
representing a permutation to be completed from partial
information~\cite[Problem~4.1]{Hanika2015}.  For a finite set $V$ and linear orders
$L_1,L_2$ on $V$,
label the elements of $V$ by
$1,\ldots,|V|$ in their $L_1$ order.  Reading these labels in their $L_2$
order gives the induced permutation $\Pi(V,L_1,L_2)$.  We use the following
formulation of the problem.

\begin{definition}[Forbidden permutation CSP]
Fix a finite set $B$ of permutations.  The problem $\FPCSP(B)$ takes as input
a finite ground set $V$ and two acyclic directed relations $E_1,E_2$ on $V$.
It asks whether there are linear orders $L_1\supseteq E_1$ and
$L_2\supseteq E_2$ such that
\[
  \Pi(V,L_1,L_2)\in\Av(B).
\]
\end{definition}

Insert-maximum extension is the restriction in which the first order is
fixed and the second order already specifies the relative order of the root
entries.

\begin{proposition}
\label{ext:prop:extendmax-to-fpcsp}
For every fixed finite forbidden set $B$, there is a polynomial-time many-one
reduction from $\ExtendMax_B^{\mathrm{un}}$ to $\FPCSP(B)$.
\end{proposition}

\begin{proof}
Let $(\rho,1^k)$ be an instance of $\ExtendMax_B^{\mathrm{un}}$, where
$\rho\in\mathfrak S_n$, and put $N=n+k$.  Take
$V=\{1,\ldots,N\}$ and define
\[
  E_1=\{(i,i+1):1\leq i<N\}
\]
and
\[
  E_2=\{(\rho_i,\rho_{i+1}):1\leq i<n\}.
\]
Every linear order extending $E_1$ is the natural order on $V$.  A linear
order extends $E_2$ exactly when its restriction to $\{1,\ldots,n\}$ lists
those elements in the order
$\rho_1,\rho_2,\ldots,\rho_n$.  Consequently, the permutations induced by
linear-order extensions of $(E_1,E_2)$ are exactly the permutations of
length $N$ obtained from $\rho$ by inserting the new maximum entries
$n+1,\ldots,N$.

It follows that $(\rho,1^k)$ is a Yes-instance of
$\ExtendMax_B^{\mathrm{un}}$ exactly when $(V,E_1,E_2)$ is a Yes-instance of
$\FPCSP(B)$.  The construction has size $O(n+k)$ and is polynomial because
$k$ is encoded in unary.
\end{proof}

\begin{corollary}
\label{ext:cor:hanika-fpcsp-np-complete}
The problem $\FPCSP(\Bfinal)$ is \NPcomplete{}.
\end{corollary}

\begin{proof}
Theorem~\ref{thm:main-hardness} and
Proposition~\ref{ext:prop:extendmax-to-fpcsp} give \NPhardness{}.  Membership in
$\mathrm{NP}$ holds for every fixed forbidden set~\cite[Remark~4.3]{Hanika2015}.
\end{proof}

Hanika asked whether $\FPCSP(B)$ is \NPcomplete{} for some fixed set of
patterns and noted that no \NPhard{} example was known.
Corollary~\ref{ext:cor:hanika-fpcsp-np-complete} answers this question with
the explicit fixed set $B=\Bfinal$.


\section{Marker and Constraint Permutations}
\label{ext:app:coordinates}
\label{ext:app:fixed-permutation-data}

The four marker permutations of Definition~\ref{ext:def:markers} and the four
constraint permutations of Definition~\ref{ext:def:constraint-permutations} are
\[
\begin{aligned}
D   &= 9,3,10,8,6,4,2,5,1,7,\\
S   &= 4,9,7,1,6,10,5,3,8,2,\\
E   &= 5,9,3,8,10,7,1,6,4,2,\\
N   &= 8,1,7,5,10,4,9,3,6,2,\\[1mm]
Q_L &= 8,10,7,5,3,9,2,6,1,4,\\
Q_R &= 3,5,10,8,6,4,2,9,1,7,\\
C_L &= 5,10,2,8,6,4,7,3,1,9,\\
C_R &= 8,10,7,4,6,1,5,3,9,2.
\end{aligned}
\]

\begin{figure}[ht]
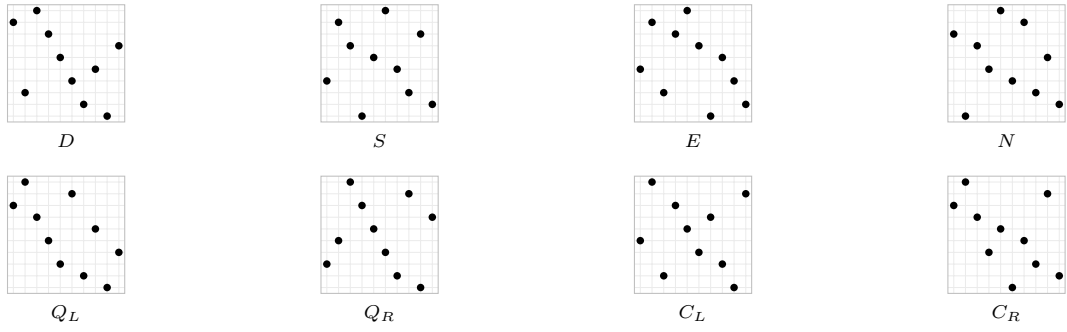

\centering
\newcommand{\mk}[2]{\begin{tabular}{@{}c@{}}
  \begin{permpic}[0.155]{10}\pts{#2}\end{permpic}\\[0.5mm] {\scriptsize $#1$}\end{tabular}}
\mk{D}{9,3,10,8,6,4,2,5,1,7}\hfill
\mk{S}{4,9,7,1,6,10,5,3,8,2}\hfill
\mk{E}{5,9,3,8,10,7,1,6,4,2}\hfill
\mk{N}{8,1,7,5,10,4,9,3,6,2}

\medskip
\mk{Q_L}{8,10,7,5,3,9,2,6,1,4}\hfill
\mk{Q_R}{3,5,10,8,6,4,2,9,1,7}\hfill
\mk{C_L}{5,10,2,8,6,4,7,3,1,9}\hfill
\mk{C_R}{8,10,7,4,6,1,5,3,9,2}
\caption{The eight simple permutations.  Their pairwise distinctness,
simplicity, and longest decreasing subsequence length are given in
Propositions~\ref{ext:prop:marker-facts}
and~\ref{ext:prop:constraint-permutation-facts}.}
\label{ext:fig:fixed-permutations}
\end{figure}

\end{document}